\documentclass{aa501}

\usepackage{graphicx}

\begin{document}
    \title{Intermediate resolution H$\beta$ spectroscopy and photometric 
monitoring of 3C 390.3}
    \subtitle{I. Further evidence of a nuclear accretion disk \thanks{based on observations carried out at SAO RAS, Guillermo Haro, Abastumani and Sternberg Observatories.}}

\author{A.I.\,Shapovalova\inst{1}
	\and A.N.\,Burenkov\inst{1}
	\and L.\,Carrasco\inst{2,3}
	\and V.H.\, Chavushyan\inst{2}
	\and V.T.\,Doroshenko\inst{4}
	\and A.M.\,Dumont\inst{5}
	\and V.M.\,Lyuty\inst{4}
	\and J.R.\,Vald{\'e}s\inst{2}
	\and V.V.\,Vlasuyk\inst{1}
	\and N.G.\,Bochkarev\inst{4} 
	\and S.\,Collin\inst{5}
	\and F.\,Legrand\inst{2}
	\and V.P.\,Mikhailov\inst{1}\and O.I.\,Spiridonova\inst{1}
	\and O.\,Kurtanidze\inst{6}\and M.G.\,Nikolashvili\inst{6}}
	\institute{Special Astrophysical Observatory of the Russian AS, Nizhnij Arkhyz,
Karachaevo-Cherkesia, 369167, Russia\\
	\email {ashap@sao.ru}
	\and
	Instituto Nacional de Astrof\'{\i}sica, Optica y Electr\'onica,
	Apartado Postal 51, C.P. 72000, Puebla, Pue., M\'exico\\
	\email {carrasco@inaoep.mx}
	\and
	Observatorio Astron\'omico Nacional, UNAM, Apartado Postal 877, C.P. 22860, Ensenada B.C., M\'exico
	\and
	Sternberg Astronomical Institute, University of Moscow, Universitetskij Prospect 13, Moscow 119899, Rusia
	\and
	DAEC, Observatorie de Paris, section de Meudon, Place Jansen, F-92195 Meudon, France
	\and	
 Abastumani Astrophysical Observatory, Georgian AS, Mt. Kanobili, 383762, Abastumani, Georgia}
\offprints{A.I.~Shapovalova  or L.~Carrasco}

\date  {Received: February 21, 2001 / Accepted: June 13, 2001}
\titlerunning {H$\beta$ monitoring of 3C 390.3}
\authorrunning{Shapovalova, A.I., Burenkov, A.N., Carrasco, L., et al.}

\abstract{ 
We have monitored the AGN 3C~390.3 between 1995 and 2000.
A historical B-Band light curve dating back to 1966, shows a large
increase in brightness during 1970 -- 1971, followed by a gradual decrease
down to a minimum in 1982. During the 1995--2000 lapse the broad H$\beta$ emission
and the continuum flux varied by a factor of
$\approx $ 3. Two large amplitude 
outbursts, of different duration, in continuum and $\rm H\beta$ light were 
observed ie.: in October 1994 a brighter flare that lasted  $\approx $1000 days
and in July 1997 another one that lasted $\approx$ 700 days
were detected. The response time lag of the emission lines relative to flux changes of the continuum, has been found to vary with time ie. during 1995--1997 a lag of about 100 days is evident, while during 1998--1999 a double valued lag of $\approx $ 100 days and $\approx $ 35 days is present in our data.
The flux in the H$\beta$ wings and line core vary simultaneously, a behavior indicative of predominantly circular motions in the BLR.
Important changes of the H$\beta$ emission profiles were
detected: at times, we found profiles with prominent asymmetric wings, as those normaly 
seen in Sy1s, while at other times, we observe profiles with weak almost symmetrical
wings, similar to those of Sy1.8s. We further dismiss the hypothesis that the double 
peaked $\rm H\beta$ profiles in this object originate in a massive binary BH. Instead,  
we found that the radial velocity difference between the red and blue bumps is 
anticorrelated with the light curves of H$\beta$ and continuum radiation. This implies that the zone that contributes with most of the energy to the emitted line, changes in radius
within the disk. The velocity difference increases,
corresponding to smaller radii, as the continuum flux decreases. When 
the continuum flux increases the hump velocity difference decreases. These 
transient
phenomena are expected to result
from the variable accretion rate close to the central source.
The optical continuum and the $\rm H\beta$ flux variations might be
related to changes in X-ray emission modulated by a variable
accretion rate, changing the surface temperature of the disk,
as a result of a variable X-ray irradiation ( Ulrich, \cite{ulr}).
Theoretical $\rm H\beta$ profiles were computed for an accretion disk,
the observed profiles are best reproduced by an inclined disk ($25\degr$)
whose region of maximum emission is located roughly at 200 $R_g$.
The mass of the black hole in
3C 390.3, estimated from the reverberation analysis is
$M_{rev}\approx 2.1\times 10^9\,M_{\sun}$, 5 times larger than
previous estimates (Wandel et al.\cite{wand}).
}
\maketitle

\section{Introduction}
Quasars and Active Galactic Nuclei (AGNs) are amongst the most
luminous objects in the Universe and provide a unique test case
for several processes of astrophysical importance. A relevant
characteristic of some of these objects is a variable broad
emission line spectrum. The region where broad lines are formed
(hereinafter BLR) is rather close to the central engine
and may hold basic information to the understanding of the formation and
fueling of AGNs and quasars, a process not yet fully understood.

Long temporal baseline spectral monitoring of the nuclei of some
AGNs has revealed a time lag in the response of the broad emission
lines relative to flux changes of the continuum. This lag depends on
the size, geometry and physical conditions of the emitting
region. Thus, the search for correlations between nuclear
continuum changes and flux variations of the broad emission
lines may serve as a tool for mapping the geometrical and
dynamical structure of the BLR (see Peterson, \cite{pet3}\,
and references therein). From the study of the responses of lines of
different ions, to changes of the nuclear
continuum and the comparison of these data with photoionization model
predictions, one could infer some of the physical conditions in
the BLR as a function of the distance from the central source
(Dumont {\rm et al.}\, \cite{dum3}).

From changes of the broad emission line profiles, we could
distinguish, in principle, different kinematical models for the BLRs,
including cases in which matter is undergoing accelerated motions,
ie. falling or outflowing, keplerian rotation in the gravitational
field of the central body, and some others (Bochkarev \& Antokhin \cite{boch1};
Blanford \& McKee \cite{blan}; Antokhin \& Bochkarev \cite{antokhin}).

During the last decade, the study of the BLRs in some objects
has met considerable success, mainly due to an increasing
number of coordinated multiwavelength monitoring campaigns
through the international ``AGN Watch'' program. This program
has provided data with good temporal frequency and coverage for
a number of selected Seyfert galaxies: NGC 5548 (Clavel {\rm et al.}\,\cite{clav2};
Peterson {\rm et al.}\,\cite{pet2},\cite{pet4},
\cite{pet5}, \cite{pet7}; Maoz {\rm et al.}\,\cite{maoz2};
Dietrich {\rm et al.}\, \cite{die1}; Korista {\rm et al.}\,\cite{kor};
Chiang {\rm et al.}\,\cite{chiang}); NGC 3783 (Reichert {\rm et al.}\,\cite{reich};
Stirpe {\rm et al.}\, \cite{stirpe}; Alloin {\rm et al.}\,\cite{alloin}); NGC 4151
(Crenshaw {\rm et al.}\,\cite{crenshaw};
Kaspi {\rm et al.}\,\cite{kaspi}; Warwick {\rm et al.}\, \cite{war}; Edelson
{\rm et al.}\,\cite{edelson}); Fairall 9 (Rodriguez-Pascual
{\rm et al.}\,\cite{rodri}; Santos-Lleo {\rm et al.}\, \cite{santos});
NGC7469 (Wanders {\rm et al.}\,\cite{wan};
Collier {\rm et al.}\,\cite{collier}) and 3C 390.3 (Leighly {\rm et al.}\,
\cite{lei2}; Dietrich {\rm et al.}\, \cite{die2}; O'Brien {\rm et al.}\,
\cite{obri2}). It has been found that a significant part of the BLR
variability can be associated with a region located within some light
weeks from the central source. The infered BLR extent is
comparable to the size of an hypothetical accretion disk
surrounding a supermassive black hole. Therefore, at least part
of the flux from the BLR arises, apparently, from the accretion
disk itself (Laor \& Netzer \cite{laor}, Dumont \& Collin-Souffrin \cite{dum1},
\cite{dum2};
Zheng {\rm et al.}\,\cite{zhe1}; Hubeny {\rm et al.}\, \cite{hub}). It is also
found that, the higher ionization lines respond faster to continuum flux changes
than those of lower ionization and that the optical and ultraviolet continua vary
without a significant time lag between them.

The results of the studies of the velocity fields in the central sources are
still ambiguous. It is not clearly established yet, whether or not, the time lags
in the response of the blue and red wings and the core of the line profile
with respect to continuum variations are the same. So far, studies of the velocity
dependent response of the \ion{C}{iv} $\lambda$1549 emission line in
NGC 5548 have found no evidence of significant radial motions in its
BLR. (Korista {\rm et al.}\, \cite{kor}).

Most of the objects, included in the multiwavelength monitoring
AGN Watch, are radio-quiet Sy1 galaxies and only one, 3C 390.3
(z=0.0561), is a well known broad line radiogalaxy. This type of
object represents about 10\% of the radio-loud AGNs. It
is also a powerful double lobed FRII radio-galaxy with a
relatively strong compact core. The two extended lobes, at a
position angle of 144$^{\circ}$, each one with a hot spot, are
separated by about 223\arcsec (Leahy \& Perley \cite{lea1}). A faint
well-collimated thin jet at  P.A. = 37$^{\circ}$, connecting the
core to the northern lobe, has been observed by Leahy \& Perley
(\cite{lea2}). The VLBI observations at 5 GHz show evidence of
superluminal motion (with v/c $\sim $4) in this parsec scale jet
(Alef {\rm et al.}\, \cite{Alef}, \cite{alef}).

The strong variability of this object, both in the continuum light and the
emission lines is well known (Barr {\rm et al.}\,\cite{bar1}; Yee \& Oke\,
\cite{yee}; Netzer \cite{netzer}; Barr {\rm et al.}\, \cite{bar2};
Penston \& Perez\, \cite{pen2}; Clavel \& Wamsteker\,\cite{clav1};
 Veilleux \& Zheng\, \cite{veilleux}; Shapovalova {\rm et al.}\, 
\cite{shap};
Zheng \cite{zhe3}; Wamsteker {\rm et al.}\, \cite{wam2};
Dietrich {\rm et al.}\,\cite{die2}; O'Brien {\rm et al.}\,\cite{obri2}). The
object is also a highly variable X-ray source, with a spectrum
showing a broad Fe K$\alpha$ line (Inda {\rm et al.}\, \cite{inda}; 
Eracleous {\rm et al.}\, \cite{erac2}; Wozniak {\rm et al.}\, \cite{woz}). 
During a multiwavelength monitoring campaign in 1995, several large-amplitude
X-ray flares were observed in 3C 390.3; in one of them the X-ray flux increased 
by a factor of 3 in 12 days (Leighly {\rm et al.}\,\cite{lei2}). Evenmore,
Leighly \& O'Brien (\cite{lei3}) have presented evidence for
nonlinear X-ray variability.

In an analysis of IUE spectra of 3C 390.3, obtained during the
1978--1986 period, Clavel \& Wamsteker \cite{clav1} detected a
variability time lag of 50 and 60 days between the broad emission lines 
of \ion{C}{iv}
$\lambda$1549 and Ly$\alpha$, and the
UV-continuum. However, from a reanalysis of the same
spectral data, Wamsteker {\rm et al.} \cite{wam2} have derived a lag
of 116 $\pm$ 60 days for \ion{C}{iv} and 143 $\pm$ 60 days for Ly$\alpha$.
Furthermore, from IUE monitoring data for December 1994 to
March 1996, O'Brien et al. (\cite{obri2}) obtained a corresponding
lag of 35 days for the \ion{C}{iv} $\lambda1549$ emission line and 60
days for Ly$\alpha$. Yet, from optical monitoring of 3C 390.3 in the
1994--1995 period, Dietrich {\rm et al.} (\cite{die2}) derived a time
lag of about 20 days for the Balmer lines response to changes in the X-ray 
continuum. Furthermore, at that time, no
temporal lag between the optical and the UV or X-ray continua
changes was detected. The UV-bump, usually observed in a large
number of Seyfert galaxies is very weak or even absent in 3C 390.3
(Wamsteker {\rm et al.} \,\cite{wam2}). Our object is a prototype
of a class that shows very broad, double peaked emission line profiles, 
to explain this profile, a number of possible scenarios have been 
suggested. These models contemplate the following hypothetical physical 
models:

\begin{enumerate}
\item  Supermassive binary black holes (Gaskel\, \cite{gas1}). If each
of the black holes (BH) had its own associated BLR, then the blue and
red bumps, seen in the broad lines, could be assesed to the
corresponding BH. As a result of precession of the binary BH,
secular displacements of the peak wavelengths are expected. From 
the analysis of 3C 390.3 spectra, obtained during 20 years (1968--1988), 
Gaskel (\cite{gas4}) detected a displacement of
the blue bump in H$\beta$ at $+1500$\,km/s, and determined from
the radial velocity changes a possible period of about 300
years that corresponds to 7$\times$10$^9$ M$_{\sun}$ for the BHs.
However, Eracleous {\rm et al.} (\cite{erac3}), having added data on the
displacement of the H$\alpha$ blue peak for the 1988-1996 lapse,
obtained a period for the binary BH, P $>$ 800 years, with an
estimated total mass for the binary BH larger than 10$^{11}$
M$_{\sun}$. These authors argued that such a large value for the mass
of the BH is difficult to reconcile with the fact that, the BHs found
so far in the centers of galaxies, have masses lower than 10$^{9}$ M$_{\sun}$.
Therefore, they reject the hypothesis of a binary black hole
for 3C 390.3.

\item Outflowing biconical gas streams (Zheng {\rm et al.}\, \cite{zhe1}). In 
this case the response of the broad emission
lines to variability of the continuum should occur from the blue
wing to the red wing passing through the line core, ie. the
blue and red wings must show different time lags. Although Wamsteker
{\rm et al.}\, (\cite{wam2}) claimed that the Ly$\alpha$ and \ion{C}{iv}
$\lambda$1549 blue wings lagged behind the red ones, implying
radial gas motions, this result is not confirmed
by Zheng (\cite{zhe3}). Apparently, these contradictory results are caused by 
the rather poor 
temporal sampling of the UV data ($\sim$94 days).
Dietrich {\rm et al.} (\cite{die2}) have not detected any lag in the response
of the blue and red wings relative to the core of H$\alpha$
and H$\beta$ emission. It was also found that the
flux in the wings of H$\beta$, Ly$\alpha$ and \ion{C}{iv} $\lambda$1549
vary quasi-simultaneously (Shapovalova {\rm et al.}\,\cite{shap}; O'Brien {\rm et al.}\,\cite{obri2}). 
These results imply that there
are no significant radial motions in the BLR, and Livio \& Xu (\cite{livio}) 
have shown that the double-peaked lines seen in 3C 390.3 cannot be produced 
by two line-emitting streams, since the emission region
of the far cone jet would be obscured by the optically thick accretion disk.

\item Broad Line emission from an accretion disk in 3C 390.3 has been
suggested by Perez {\rm et al.}\,(\cite{perez}). In this case, the emission
from an inclined (30$^{\circ}$) accretion disk provides a good fit
to the broad H$\alpha$ and H$\beta$ profiles. (Rokaki {\rm et al.}\, \cite{rok1}). Simple disk 
models predict two symmetrical bumps, the
blue bump being slightly brighter due to doppler-boosting effects. However, IUE data shows a blue wing of Ly$\alpha$
much brighter than the red one, this cannot be explained by a
doppler boosting effects alone. There are also epochs when the red wing
of Ly$\alpha$ is brighter than the blue one (Zheng 1996), i.e.
the flux ratio of the red to the blue wings is quite variable.
Also, quasiperiodical variations ($P \approx 10 yr$ ) of the observed flux ratios of the
blue and red wings of H$\beta$
have been reported  (Veilleux \& Zheng \cite{veilleux};
Shapovalova {\rm et al.}\, \cite{shap}; Bochkarev {\rm et al.}\,
\cite{boch2}). In order to explain these variable ratios, more complex models are 
required; for instance, hot spots
(Zheng {\rm et al.}\,\cite{zhe1}); two-arm spiral waves in the
accretion disk (Chakrabarti \& Wiita\,\cite{chakra}); or a
relativistic eccentric disk (Eracleous {\rm et al.}\,\cite{erac1}).

Long temporal base line studies of the changes in the broad emission line profile 
through systematic monitoring shall allow us to
distinguish between those models.

In this paper, we present the results of the spectral (H$\beta$) and  photometric BVRI 
monitoring of 3C 390.3 during
the 1995--1999 period. This work is part of the long-term
monitoring program for about 10 Seyfert galaxies of different
nuclear luminosities started in 1986 at the SAO RAS, and
carried out jointly, since 1998, with the INAOE (M\'exico), and several 
observatories in the Former Soviet Union (FSU) and
West-European countries (Bochkarev {\rm et al.}\, \cite{boch3}; 
Bochkarev \& Shapovalova \cite{boch4}).

\end{enumerate}

\section {Observations and Data Reduction}

\subsection{Photometry}

\begin{figure}
 \resizebox{\hsize}{!}{\includegraphics{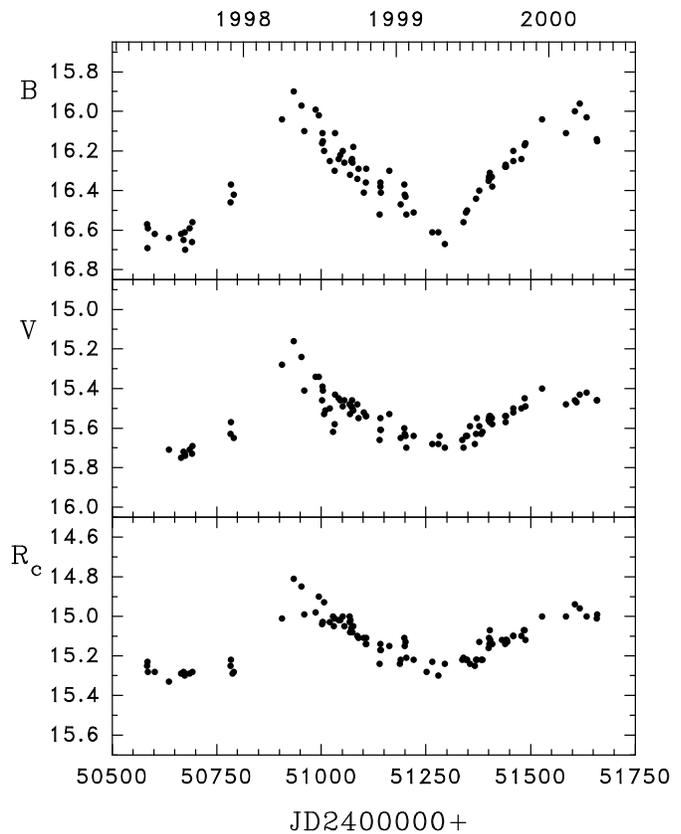}}
 \hfill
  \caption{The light curves of 3C 390.3 in the B,V and R$_\mathrm{c}$ photometric bands 
for a 10 \arcsec aperture, in the 1997--2000 period.}
  \label{bvr}
\end{figure}

\noindent The BVR photometry of 3C 390.3 was carried out
at three different observatories i.e. SAO RAS in North
Caucasus (Russia), the Crimean Laboratory of the Sternberg Astronomical Institute (CL SAI, Russia)
and the Abastumani Astrophysical Observatory (AAO) in Georgia. The observations at
the SAO RAS during 1997--2000 were obtained with the 1\,m and 60\,cm Zeiss telescopes equipped with an offset guided automatic photometer. The instrument
has a liquid nitrogen cooled CCD camera with a format of 1040 $\times$ 1160 pixels, (Amirkhanian {\rm et al.}\, 2000). The scale at the CCD being 0.45\arcsec per pixel, with a corresponding field of view of 7.5 $\times$ 8.5 arcmin.
Exposures of the morning and evening sky were adopted as
flat-field frames. Bias and dark current frames were also
obtained. Data reduction was carried out with the software
developed at SAO by Vlasyuk (\cite{vlasyuk}). The photometry is obtained by
signal integration in concentric circular apertures, of increasing size,
centered at the baricenter of the measured object. The instrumental
photometric system of this instrument is similar to those of Johnson
in B and V, and of Cousins (\cite{cousins}) in R and I. The VRI photometry during 1998--1999 was obtained at the Crimean Laboratory of the Sternberg Astronomical Institute
with the 60\,cm telescope (Zeiss-600) equipped with an ST-6 thermoelectrically cooled CCD camera. The photometric system of CL~SAI is similar to that of Johnson. Further details about SAO and Crimean observations can be found in Doroshenko {\rm et al.}\, (\cite{doroshenko}). BVRI observations at the AAO in Georgia during  1997--1998  were obtained with the 70\,cm menisk telescope equipped with a CCD camera.
These data were reduced with the IRAF package (DAOPHOT).

\noindent As local photometric standard, we adopted stars ``B'' and ``D''
of Penston {\rm et al.}\, (\cite{pen1}) which are close to 3C~390.3 in our CCD images. consequently, effects due to differential air mass are negligible for internal calibration purposes. The adopted BVRI magnitudes for the reference stars are given in Table $\ref{std}$.

\begin{table}[hbtp]
\begin{center}
\caption{\footnotesize{Adopted Photometry for the Standard Stars}}
\vspace{0.1cm}
\label{std}
\begin{tabular}{ccccccc}
\hline
    Star& $\mathrm{B}^a$ & $\mathrm{V}^a$ & $\mathrm{R}^b$ & $\mathrm{I}^b$ \\
\hline
     B  & 15.04 & 14.28 & 14.13 & 13.59 \\
     D  & 15.40 & 14.65 & 14.42 & 13.90 \\
\hline
\end{tabular}
\end{center}

$^a$From Penston {\rm et al.} (\cite{pen1}). \\
$^b$From Dietrich {\rm et al.} (\cite{die2}).\\
\end{table}

\noindent The results of the broad-band photometry of 3C 390.3
in the BV$\mathrm{R}_\mathrm{C}$ filters for a circular aperture of 10\arcsec
obtained at AAO and CL~SAI were transformed to the
SAO BVR photometric system. Our results are presented in graphical form in Figure\,$\ref{bvr}$.
\noindent The estimated mean errors in the BVRI
photometry for the entire data set are 0.026, 0.021, 0.020, 0.022 magnitudes, respectively.
In Table $\ref{table2}$ our photometric results and the
corresponding errors are listed. The last column lists, the observatory where the data was obtained i.e. (SAO, SAI --- Crimean Laboratory, Abast. - Abastumani Astrophysical Observatory in Georgia).

\begin{figure*}
\sidecaption
\includegraphics[width=12cm]{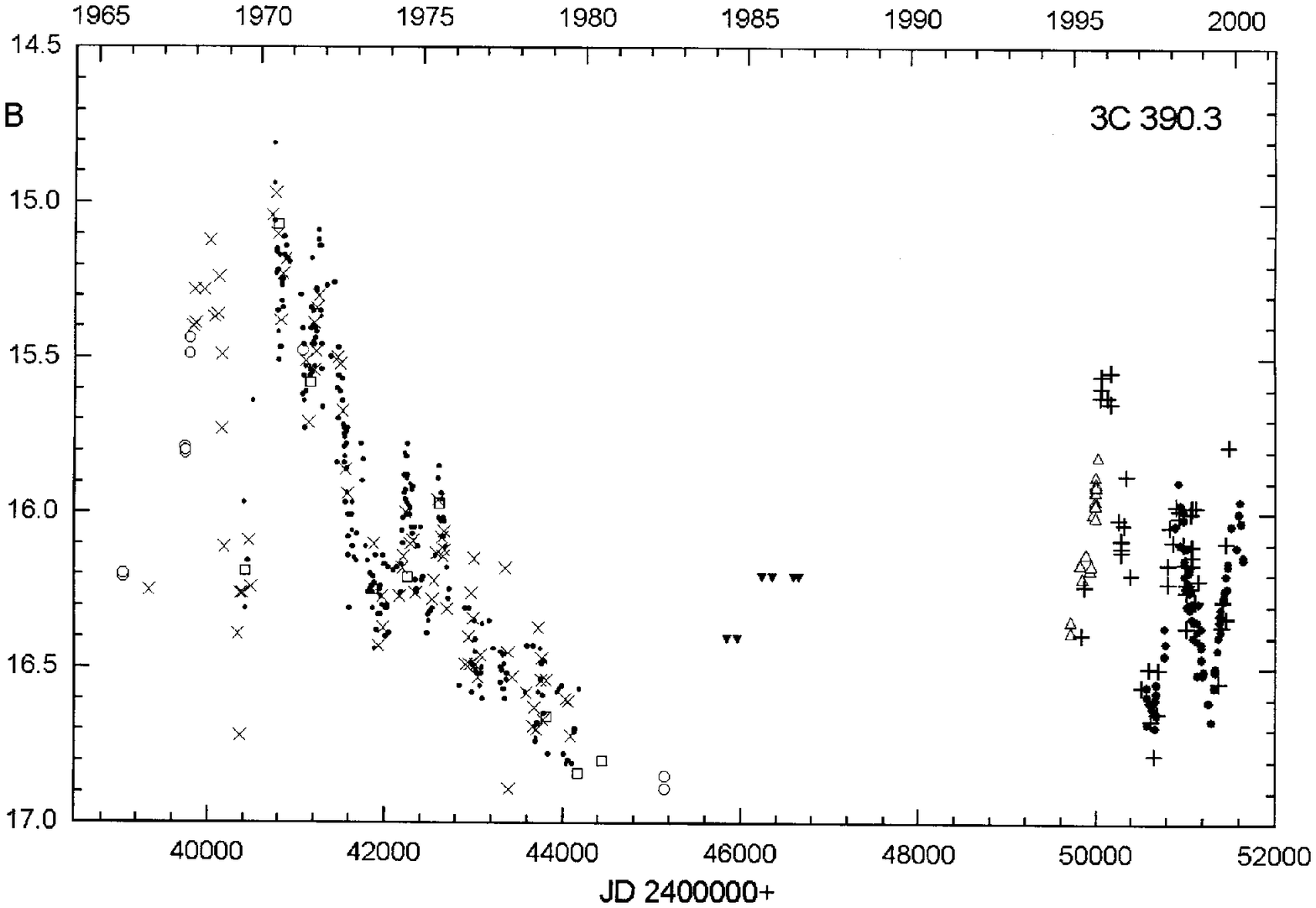}
\caption{A historical light curve for 3C 390.3.
Open cicles -- Sandage (\cite{sandage}), Neizvestny (\cite{niez});
large filled circles --  photometry (present paper);
open squares -- Yee \& Oke (\cite{yee});
small filled circles -- Babadzhanyants {\rm et al.} (\cite{bab1}, \cite{bab2}, \cite{bab3}, \cite{bab4}, \cite{bab5});
crosses -- Cannon {\rm et al.} (\cite{cannon}), Scott {\rm et al.} (\cite{scott}),
Selmes {\rm et al.} (\cite{selmes}), Pica {\rm et al.} (\cite{pica});
plusses -- spectral continuum (present paper);
open triangles -- Dietrich {\rm et al.} (\cite{die2});
filled triangles -- spectral cont. from Perez {\rm et al.} (\cite{perez})
and Lawrence {\rm et al.} (\cite{law}).}
	\label{slc}
\end{figure*}

\begin{table}
\centering
	\caption {Photometric Results}
  \title{
	\thanks {Table 2 is only available in electronic form at the
CDS via anonymous ftp to cdsarc.u-strasbg.fr (130.79.125.5) or
via http://cdsweb.u-strasbg.fr/Abstract.html}}
\label{table2}
\end{table}

\subsection {Optical Spectroscopy}
\subsubsection {Observations}

Spectra of 3C 390.3 were obtained with the 6 m and 1 m
telescopes of the SAO RAS (Russia, 1995--1999) and at INAOE's
2.1 m telescope of the ``Guillermo Haro Observatory'' (GHO) at Cananea,
Sonora, M\'exico (1998--1999). The spectra were obtained with long slit
spectrographs, equipped with CCD detector arrays. The typical
wavelength interval covered is that from 4000\,\AA\, to 7500\,\AA. The
spectral resolution varied between 4 and 15 \AA. Spectrophotometric
standard stars were observed every night. The specific information
about the instrumental set-ups in different telescopes is listed
in Table $\ref{telesc}$, there we list: 1 -- telescope; 2 -- type of
focus; 3 -- spectrograph; 4 -- CCD format; 5 -- set-up code.
The log of spectroscopic observations is given in Table $\ref{table4}$,
listed in Cols. 1 -- UT date; 2 -- Julian date; 3 -- code according
to Table 3; 4 --  projected spectrograph entrance
apertures (the first dimension being the slit-width, and
the second one the slit-length); 5 --
wavelength range covered; 6 -- spectral resolution; 7 -- seeing; 8 --  signal to noise ratio in a region of the continuum (5370--5420\AA), where there are no  prominent emission or absorption lines.
The spectrophotometric data reduction was carried out either with software developed at SAO RAS by Vlasyuk(1993) or with the IRAF package for the spectra obtained in M\'exico. The image reduction process included bias,
flat-field corrections, cosmic ray removal, 2D wavelength
linearization, background subtraction, stacking of the spectra for
every set up, and flux calibration.

\begin{table}[hbtp]
\begin{center}
\caption{\footnotesize{Characteristics of the telescopes and spectrographs}}
\vspace{0.1cm}
\label{telesc}
\begin{tabular}{ccccc}
\hline
\hline
Telescope & Focus &Equipm.&CCD& Code\\
          &       &       &(pixels)& \\
\hline
1 & 2 & 3 & 4 & 5 \\
\hline
6m SAO   & Prime      & UAGS           & 530$\times$580   & G \\
6m SAO   & Prime      & UAGS           &1024$\times$1024  & P \\
6m SAO   & Prime      & MPFS$^{\rm a}$ & 530$\times$580   & T \\
6m SAO   & Nasmyth    & Long slit      &1024$\times$1024  & N \\
1m Zeiss & Cassegrain & UAGS           & 530$\times$580   &Z1 \\
1m Zeiss & Cassegrain & UAGS           &1040$\times$1170  &Z2 \\
2.1m GHO & Cassegrain & B\&Ch          &1024$\times$1024  & M \\
\hline
\end{tabular}
\end{center}
\vspace{0.1cm}
\hspace{0.2cm}

\begin{list}{}{}
\item[$^{\rm a}$] MPFS refers to a Multi-Pupil Field Spectrograph
\end{list}

\end{table}

 
\begin{table}
\centering
	\caption{Log of the Spectroscopic Observations}
  \title{
	\thanks {Table 4 is only available in electronic form at the
CDS via anonymous ftp to cdsarc.u-strasbg.fr (130.79.125.5) or
via http://cdsweb.u-strasbg.fr/Abstract.html}}
\label{table4}
\end{table}

\begin{figure}
\resizebox{\hsize}{!}{\includegraphics{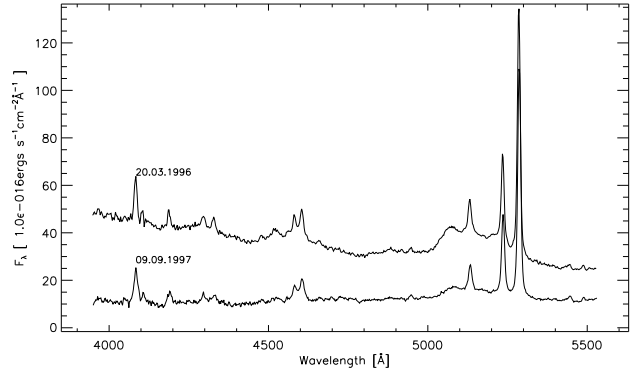}}
\caption{The spectra corresponding to the high activity (top) and the low 
activity (bottom) states}
\label{lh}
\end{figure}

\subsubsection {Absolute Calibration of the Spectra}

Since, even under good photometric conditions, the accuracy of
spectrophotometric measurements is rarely better than 10\%, for the study 
of AGN's variability the standard scheme for flux calibration by means of comparison 
with stars of 
known spectral energy distribution is not acceptable.
Instead, we use the flux of the narrow forbidden emission lines which in AGNs are non-variable in timescales of tens of
years. This is largely due to the extension of the narrow line
emitting region (NLR) and to the lower gas density present in it.
Thus, the effects of light traveling through this region,
jointly with the long recombination times (t$_\mathrm{rec}$)$\sim$100
years, for $\mathrm{n}_\mathrm{e}\sim$10$^{3}$ cm$^{-3}$), damp out the short
time scale variability. Hence, the bright narrow emission lines are
usually adopted as internal calibration for scaling AGN spectra
(Peterson 1993). However, in the case of \object{3C~390.3} there has been
some discussion about the variability of these lines. Let us address
this question in more detail.

From IUE data, Clavel \& Wamsteker (\cite{clav1}) found 
that the flux in the narrow components of Ly$\alpha$
and \ion{C}{iv} $\lambda$1549 emission lines in 3C 390.3 decreased
continuously by $\approx$ 40\% between 1978 and 1986. The involved
timescales suggest an upper limit of about 10 light years for the
size of the regions in which these lines are formed.

Zheng {\rm et al.}\, (\cite{zhe2}) reported a considerable variability of the
[\ion{O}{iii}] $\lambda$4959 narrow emission line fluxes ($\sim $1.8
times) from observations carried out between 1974 and 1990. We
believe that this result is somewhat doubtful in the light of
the following considerations: the variability is only infered when
comparing results from observations carried out with different
instrumental set-ups. They include, on the one hand, fluxes
from image dissector scanner (IDS) data, obtained with the Shane
3\,m telescope at Lick Observatory through an aperture of $2.8
\times 4$\arcsec (1974--1983 lapse) and on the other hand,
observations carried out at La Palma with the 2.5\,m Isaac Newton
telescope
(INT), equipped with an image photon counting system (IPCS)
through a 1.6\arcsec\, wide slit  (1984--1989). The lowest flux values
largely correspond to those observed at La Palma through this rather
narrow slit. Hence, it is possible that the large differences in flux
reported for the [\ion{O}{iii}] $\lambda$4959 line could be due to
effects associated with the narrow slit used. Evenmore, in the spectra published
by Zheng {\rm et al.} (\cite{zhe2}), it is clearly seen that, during 1988 and 1989, 
the red wing of
H$\beta$ showed an important enhancement and its contribution
could have also affected the determination of the [\ion{O}{iii}] line
flux. Unfortunately, these authors do not provide a detailed
discussion of their adopted procedure for flux determination.

Relevant in the context of forbidden line variability are the
observations by Yee \& Oke (\cite{yee}) that covered the 1969--1980
lapse, these were carried out with the Hale 5\,m telescope using a
multichannel spectrophotometer through an aperture of $\sim$10\arcsec.
They found that the [\ion{O}{iii}] lines and the narrow Balmer line fluxes 
did not changed during those years. Furthermore, a detailed study of the [\ion{O}{iii}] 
$\lambda$5007 flux variability in 3C 390.3
was carried out by Dietrich {\rm et al.} (\cite{die2}) during 1994--1995. The spectra 
were obtained through wide apertures (5\arcsec$\times$7.5\arcsec and 
4\arcsec$\times$10\arcsec) and were calibrated by comparison with broad band photometry. 
It was found, as expected, that the [\ion{O}{iii}] $\lambda$5007 line flux remained 
constant
(within 2.6\%)
during the observed time interval. Therefore, we consider that in the case of 3C 390.3 
there is no reliable evidence of the [\ion{O}{iii}] $\lambda\lambda $4959+5007 flux 
variability in
time scales of months to few years. Our spectra for the H$\beta$ region were scaled by
the [\ion{O}{iii}] $\lambda\lambda$4959+5007 integrated line flux under the assumption
that the latter did not change during the time interval covered by our observations 
(1995--1999). 
A value of 1.7$\times$10$^{-13}$ ergs s$^{-1}$cm$^{-2}$ (Veilleux \& Zheng \cite{veilleux}) 
for the integrated [\ion{O}{iii}] line flux was adopted. In order to calculate a 
normalization coefficient, the continuum was determined in two 30\,\AA\, wide 
clean --line free-- windows centered at 4800\,\AA\, and 5420\,\AA respectively. 
After continuum subtraction, blend separation of the H$\beta$ and [\ion{O}{iii}] 
components was carried out by means of a Gaussian fitting procedure, applied to 
the following: H$\beta$ --- broad blue, broad red and central narrow; [\ion{O}{iii}] 
$\lambda\lambda$4959,5007 --- broad and narrow components. The forbidden lines are 
represented by two Gaussian curves with an intensity ratio
I(5007)/I(4959)=2.96.

\subsubsection {Intercalibration of the Spectral Data} 
The flux of H$\beta$ was determined from scaled
spectra after continuum subtraction and removal of the [\ion{O}{iii}]
doublet. The narrow component of
H$\beta$ was not subtracted due to possible ambiguities in the
fitting scheme for the broad and narrow components of
the line. We should mention, that the contribution of
the narrow component to the integral H$\beta$ flux is only about 7\%.
As Peterson \& Collins (\cite{pet1}) pointed out, it is necessary to correct 
the fluxes for aperture effects, due to the fact that
while the BLR and nonstellar continuum are effectively pointlike
sources ($<$1\arcsec), the NLR is an extended one
($>$2--15\arcsec). Then, the measured NLR flux depends on
the size of the spectrograph's entrance aperture, and in the 
case of direct images, on the size of the adopted aperture. The light
contribution of the host galaxy to the continuum depends
also on the aperture size. Peterson {\rm et al.} (\cite{pet6}) have
shown that
the uncertainties in CCD photometry can be minimized, when
sufficiently large apertures are adopted. For an aperture of 5\arcsec$\times$7.5\arcsec,
the expected photometric errors are typically 2 to 3\,\%.

\noindent  The NLR in 3C 390.3 is more compact
than in most Sy1 galaxies, in narrow-band [\ion{O}{iii}] images, this object
shows a compact nuclear emission without signs of an extended structure (Baum {\rm et al.} 
\cite{baum}). The results of panoramic two-dimensional spectrophotometry 
of the nuclear region of this object show that the [\ion{O}{iii}] $\lambda$5007 
emission arises from a zone smaller than r~$<$~2\arcsec (Bochkarev {\rm et al.} 
\cite{boch2}). Furthermore, Osterbrock {\rm et al.} (\cite{oster}) obtained a 
rather low value for the
[\ion{O}{iii}] F($\lambda$4363)/F($\lambda$5007) line ratio, implying
a moderately high electron density in the NLR (several 10$^{6}$cm$^{-3}$).
Since the NLR in 3C 390.3 can be considered as a point
source, we did not applied corrections for aperture
effects to the nonstellar continuum to narrow line flux ratios or
to the broad to narrow line flux ratios, as the light losses in
the slit are similar for these components. However,
the light contribution to the continuum from the host galaxy does depend
on the aperture, and it is necessary to correct this effect. To acomplish this goal, 
we adopt the  scheme by Peterson {\rm et al.} (\cite{pet6}). Which is based on the ratio:

\begin{equation}
F_{con}~=~F(4959+5007)\left[\frac{F_{con}}{F(4959+5007)}\right]_{obs} - G,
\end{equation}

\noindent where $F(4959+5007)$ is the absolute flux in the [\ion{O}{iii}] doublet and
the value in brackets is the continuum to [\ion{O}{iii}] lines
observed flux ratio; $G$ is an aperture dependent correction factor to
account for the host galaxy light. Since most of our spectra were obtained
with the 6\,m telescope through an aperture of 2\arcsec$\times 6$\arcsec, this
instrumental setup was adopted as the standard one (i.e. $G=0$ by definition). The 
value of $G$ for the spectra obtained at the 2.1\,m telescope (aperture of $2.5\arcsec\times 6.0$\arcsec) is $G=0.046$, while the value obtained for the 1\,m telescope (aperture of $8\arcsec\times 19.8$\arcsec) is $G=0.321\pm 0.028$.
In Table~$\ref{table5}$ we list the derived values of the H$\beta$ flux in
the 4990 \AA\ to 5360 \AA\ wavelength range, jointly with the continuum flux at 5125 \AA\. 
The mean error (uncertainty) in our flux
determination for both, the H$\beta$ and the continuum is $\approx
$ 3\%. These quantities were estimated by comparing results
from spectra obtained within a time interval shorter than 2 days.

In Table $\ref{table5}$ we list our results, there we give: 1 -- The Julian date;
2 -- a code, according to Table $\ref{telesc}$; 3 -- F(H$\beta$) the total
flux (in units of 10$^{-13}$ ergs s$^{-1}$cm$^{-2}$);
4 --  $\varepsilon_{\mathrm{H}\beta} $ - the H$\beta$ flux error; 5 --  $\rm F_\mathrm{c}$ --
the continuum flux at 5125\,\AA (in units of 10$^{-15}$ ergs s$^{-1}$cm$^{-2}$A$^{-1}$), 
reduced to the 6 m telescope aperture
(2\arcsec$\times$6\arcsec); 6 -- $\varepsilon_c$ - the estimated continuum flux error.

\begin{table}[hbtp]
\begin{center}
\caption{Observed H$\beta$ and continuum fluxes}
\footnotesize
\label{table5}
\begin{tabular}{lccccc} \hline \hline
 JD & Code & F(H$\beta$)$^{\rm \star}$ & $\varepsilon_{\mathrm{H}\beta} $  & $\rm F_c^{\star\star}$ & $\varepsilon_\mathrm{c}$
\\
 2440000+&&&&&\\ \hline
  49832.424 & G  & 2.107 &  0.070 &  1.309  &  0.043 \\
  49863.375 & T  & 2.715 &  0.090 &  1.576  &  0.052 \\
  50039.156 & G  & 2.537 &  0.084 &  2.636  &  0.087 \\
  50051.143 & G  & 2.840 &  0.094 &  2.703  &  0.089 \\
  50052.149 & G  & 2.368 &  0.078 &  2.923  &  0.097 \\
  50127.602 & N  & 3.073 &  0.101 &  2.529  &  0.084 \\
  50162.580 & N  & 3.335 &  0.110 &  2.571  &  0.085 \\
  50163.553 & N  & 3.502 &  0.116 &  2.756  &  0.091 \\
  50249.542 & G  & 2.937 &  0.097 &  1.851  &  0.061 \\
  50276.567 & G  & 3.202 &  0.106 &  1.638  &  0.054 \\
  50277.556 & G  & 2.961 &  0.098 &  1.681  &  0.056 \\
  50281.434 & G  & 2.775 &  0.092 &  1.607  &  0.053 \\
  50305.489 & G  & 2.908 &  0.096 &  1.813  &  0.060 \\
  50338.319 & Z2 & 2.621 &  0.087 &  1.801  &  0.059 \\
  50390.435 & G  & 2.459 &  0.081 &  1.592  &  0.053 \\
  50511.622 & G  & 1.761 &  0.058 &  1.187  &  0.039 \\
  50599.370 & G  & 1.729 &  0.057 &  1.167  &  0.039 \\
  50618.523 & G  & 1.852 &  0.061 &  0.996  &  0.033 \\
  50656.499 & G  & 1.585 &  0.052 &  0.922  &  0.030 \\
  50691.463 & G  & 1.432 &  0.047 &  1.076  &  0.036 \\
  50701.576 & N  & 1.301 &  0.043 &  1.160  &  0.038 \\
  50808.582 & G  & 1.718 &  0.057 &  1.551  &  0.051 \\
  50813.195 & G  & 1.638 &  0.054 &  1.580  &  0.052 \\
  50835.631 & N  & 1.928 &  0.064 &  1.799  &  0.059 \\
  50867.560 & N  & 2.080 &  0.069 &  1.685  &  0.056 \\
  50904.627 & G  & 1.969 &  0.065 &  1.970  &  0.065 \\
  50940.354 & N  & 2.535 &  0.084 &  1.832  &  0.060 \\
  50942.342 & N  & 2.531 &  0.084 &  1.835  &  0.061 \\
  50990.302 & N  & 2.481 &  0.082 &  1.748  &  0.058 \\
  51010.719 & M  & 2.556 &  0.084 &  1.582  &  0.052 \\
  51019.723 & M  & 2.544 &  0.084 &  1.408  &  0.047 \\
  51023.491 & G  & 2.617 &  0.086 &  1.596  &  0.053 \\
  51025.431 & G  & 2.370 &  0.078 &  1.408  &  0.047 \\
  51055.376 & N  & 2.313 &  0.076 &  1.509  &  0.050 \\
  51074.506 & Z1 & 2.546 &  0.084 &  1.537  &  0.051 \\
  51081.429 & M  & 2.250 &  0.074 &  1.508  &  0.050 \\
  51082.429 & M  & 2.345 &  0.077 &  1.617  &  0.053 \\
  51083.429 & M  & 2.413 &  0.080 &  1.570  &  0.052 \\
  51112.259 & G  & 2.339 &  0.077 &  1.433  &  0.047 \\
  51130.169 & Z1 & 2.019 &  0.067 &  1.705  &  0.056 \\
  51372.516 & P  & 1.610 &  0.053 &  1.163  &  0.038 \\
  51410.309 & P  & 1.931 &  0.064 &  1.399  &  0.046 \\
  51426.208 & P  & 1.784 &  0.059 &  1.393  &  0.046 \\
  51454.674 & M  & 2.223 &  0.073 &  1.660  &  0.055 \\
  51455.172 & P  & 2.111 &  0.070 &  1.415  &  0.047 \\
  51491.592 & M  & 2.747 &  0.091 &  2.321  &  0.077 \\
\hline

\end{tabular}
\end{center}
\begin{list}{}{}
\item[$^{\rm \star}$] -- in units of 10$^{-13}$ ergs s$^{-1}$cm$^{-2}$
\item[$^{\rm \star\star}$] -- in units of 10$^{-15}$ ergs s$^{-1}$cm$^{-2}$A$^{-1}$
\end{list}
\end{table}

\section{Data analysis}
\subsection{A historical B light curve}

\noindent In order to construct a light curve in the B band
dating back to 1966, in addition to our photometric data for the 1997--2000 
period, we 
collected from the literature all the available photometry for 3C~390.3 reported 
during 
the last 30 years. We included the photoelectric observations, carried out in 
1965--1967, 
1971 and 1982 by Sandage (\cite{sandage}) and Neizvestny (\cite{niez});
spectrophotometric observations for 1970--1979 by Yee \& Oke (\cite{yee}),
from which we derived B magnitudes in excellent agreement with
contemporary photoelectric photometry. We also make use of the photographic 
photometry available, ie. 1967--1980 from Cannon {\rm et al.}\, (\cite{cannon}); 
Babadzhanyants {\rm et al.} (\cite{bab1}; \cite{bab2}; \cite{bab3}; \cite{bab4}; 
\cite{bab5}); 
Selmes {\rm et al.}\, (\cite{selmes}); Scott {\rm et al.} (\cite{scott}); Pica 
{\rm et al.} (\cite{pica}). In order to increase their accuracy, the photographic 
observations were 
averaged in 5 day 
bins. We also included a few spectral observations by P\'erez {\rm et al.} 
(\cite{perez}), 
Lawrence {\rm et al.} (\cite{law}) and the International Monitoring data in 
the framework of the AGN Watch consortium for 1994--1995 (Dietrich {\rm et al.} 
\cite{die2}) and spectral
continuum light curve data, discussed here. The resulting
light curve in B band is shown in Fig.~$\ref{slc}$. There, we can see a
noticeable increase in luminosity with significant fluctuations between 1965
and 1977. The amplitude of the outburst from the middle of 1969 until the
middle of 1970 was about $2^m$. In 1979 the brightness of 3C~390.3 decreased
to $17^m$ and remained at this level until 1983. At that time,
Penston \& P\'erez (\cite{pen2}) noted that the broad component of
H$\beta$ had disappeared, and the spectrum of the object became quite similar to 
that of a Sy2 galaxy. A similar behavior was observed in NGC~4151,
for which, after a long lasting photometric minimum between 1984 and 1989 the 
broad wings of H$\beta$ also disappeared and was then classified as Sy2 at that 
time (Penston \& P\'erez \cite{pen2}; Lyuty {\rm et al.} \cite{lyuty}). In the 
case of \object{3C~390.3}, after maximum light
in 1970, on the descending branch of the light curve, important oscillations
with an amplitude up to $1^m$ were observed. Unfortunately, we have not found
photometric observations of this object for the 1983--1993 period and
hence, the photometric behavior of 3C~390.3 during this
period of time is unknown. A detailed study of the character of the visible
light variability in 3C~390.3 will be presented in a separate paper.

\subsection{Variability of the H$\beta$ emission line and the
optical continuum}

The photometric data obtained during 1997--2000 is plotted in  
fig.\,$\ref{bvr}$. The light curve shows an almost sinusoidal change
in brightness with a maximum amplitude of about $0^m.8$ in the B band. Brightness 
maxima ocurred in May-June 1998 and in March-April 2000.  
The light curves in the V and R 
bands are similar in shape to those of the B band, but have a smaller amplitude ($\approx0^{m}.5$),
as expected from a blue variable continuum.
In  fig.\,$\ref{bvr}$, one can also notice small amplitude light 
fluctuations, superimposed on longer time--scale changes.
Between 1995 and 1999 the H$\beta$ flux changes reached a
maximum amplitude of 2.7, while those of the continuum at
5125\,\AA, show a maximum amplitude of $\approx$ 3.2 (Table $\ref{table5}$). 
These changes are evident in a simple inspection of the spectrum of the high-activity 
state (20 Mar 1996) as compared with corresponding one of the low-activity state (9 Sep 1997). 
These are plotted in Figure $\ref{lh}$. From the combined light curve, (Fig.$\ref{slc}$). 
For the 1995--1999 time interval, we confirm a maximum amplitude flux variations in B of 
$\approx $ 3.2. This value is in excellent agreement with the value derived from 
spectroscopic continuum fluxes, rendering further support to our basic assumption: ie. 
that the flux of the [\ion{O}{iii}] lines did not change during the period of time 
covered by our observations.


\begin{figure}
\resizebox{\hsize}{!}{\includegraphics{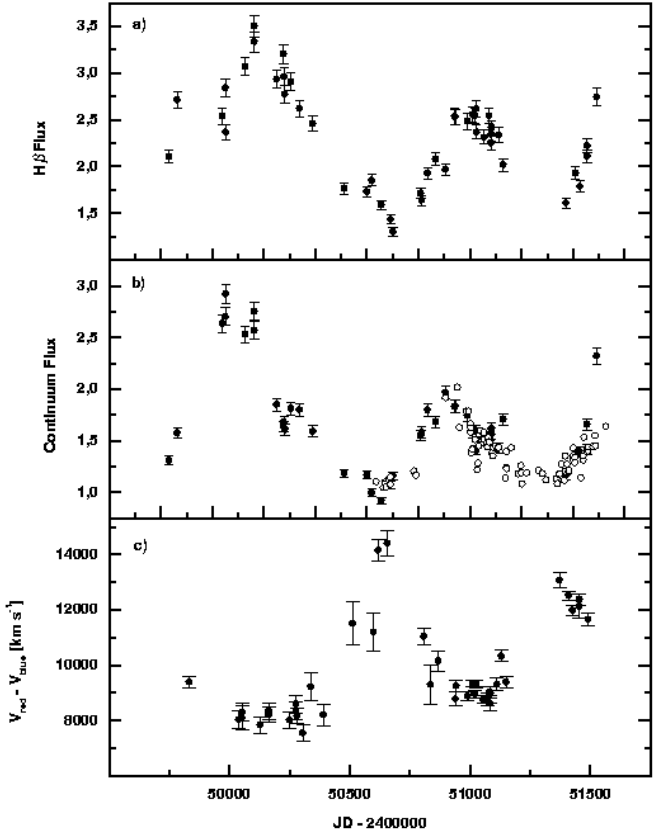}}
\caption{3C~390.3 Light curves of the H$\beta$ panel {\bf a} and the 5125\AA\ optical continuum emission 
panel {\bf b} during 1995--2000. Filled
circles represent the spectral data. The photometric points are
represented by open circles. Units are 10$^{-13}$ ergs
cm$^{-2}$s$^{-1}$ for H$\beta$ fluxes and 10$^{-15}$ ergs
cm$^{-2}$s$^{-1}$A$^{-1}$) for the continuum ones.
In panel {\bf c} we plot - V$_\mathrm{red}$ - V$_\mathrm{blue}$, the velocity difference
of the peak emission of the red and blue bumps (in km~s$^{-1}$), derived from the H$\beta$ 
difference profiles}
\label{hb-c}
\end{figure}

Figures $\ref{hb-c}${\bf a} and $\ref{hb-c}${\bf b} show the light curves for
the integrated H$\beta$ (4990--5360 \AA\ ) and continuum
(5125 \AA) fluxes, these data are given in Table 5. 
In order to improve the time resolution of our continuum
data set, we have added some photometric points in the V band to our spectral 
continuum data. The V band data
was converted into flux adopting the calibration by Johnson (\cite{john}). A 
comparison of the spectral continuum photometry at 5125\,\AA\, (F$_{c}$(5125))
with 10 simultaneous observations in the V band, yields the following 
transformation equation:

\begin{equation}
F_{c}(5125)=(0.941\pm 0.139)\times F_{\lambda}(\mathrm{V})-(0.722\pm 0.326).
\end{equation}
(correlation coefficient $r=0.923$). Here the fluxes are given in units of 10$^{-15}$ ergs~cm$^{-2}$s$^{-1}$A$^{-1}$).

\noindent In Figures $\ref{hb-c}$a and $\ref{hb-c}$b two outbursts in H$\beta$ and the
continuum are evident, with time intervals between minima of $\approx $ 
1000 days (Oct. 1994 -- Jul. 1997) and  $\approx $ 700 days (Jul. 1997 -- Jun. 1999), respectively. 
Also, the presence of a time lag between the continuum light changes 
and the response of the H$\beta$ line, is also noticeable.

\begin{figure}
\resizebox{\hsize}{!}{\includegraphics{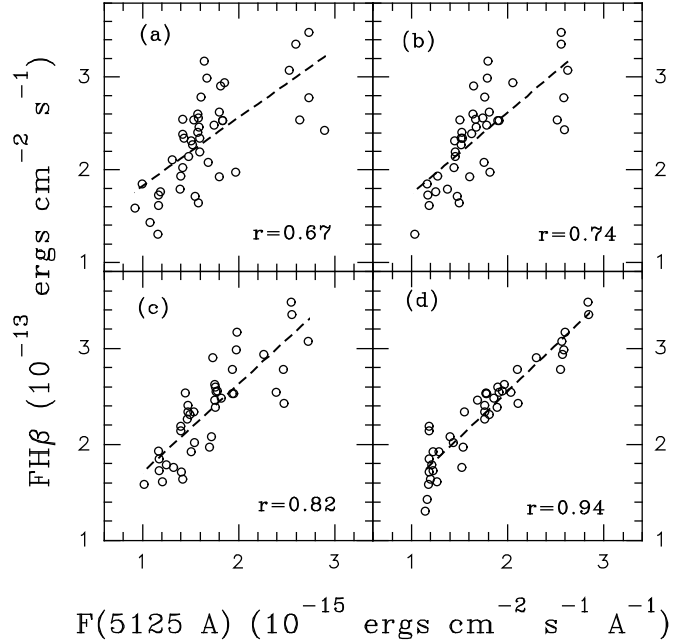}}
\caption{The F(H$\beta$) and $F_\mathrm{c}$(5125) plotted
against each other. In panels a, b, c \& d 
these quantities are plotted taking into account a time lag of 
0, 20 , 40 and 100 days, respectively. The correlation coefficients for the linear 
regressions are also given.}
\label{regr}
\end{figure}

Fig.~$\ref{regr}$a,b,c,d shows the observed correlation between 
the continuum and H$\beta$ light curves, for a time lag between the continuum 
and line variations of 0, 20, 40 and 100 days respectively. There, it is clearly 
seen that the variance of the data is
minimal while the correlation coefficient is maximal (r=0.94) for
a delay of 100 days. In latter case, we obtain the following relationship:

\begin{equation}
 F(\mathrm{H}\beta)=(0.957\pm 0.056)\times F_\mathrm{c}(5125) + (0.640\pm0.102)
\end{equation}

\noindent Here $F_\mathrm{c}$(5125) 
fluxes are expressed in units of $\rm 10^{-15}~ergs~cm^{-2}~s^{-1}~A^{-1}$ 
and the $F(\mathrm{H}\beta$) fluxes in units of $\rm 10^{-13} ergs~cm^{-2}~s^{-1}$.
The values chosen for the delays are typical values previuosly
mentioned in the literature.
In Sect. 3.4 we present a detailed discussion
of the delays from a cross--correlation analysis of our data.

\begin{figure}
\resizebox{\hsize}{!}{\includegraphics{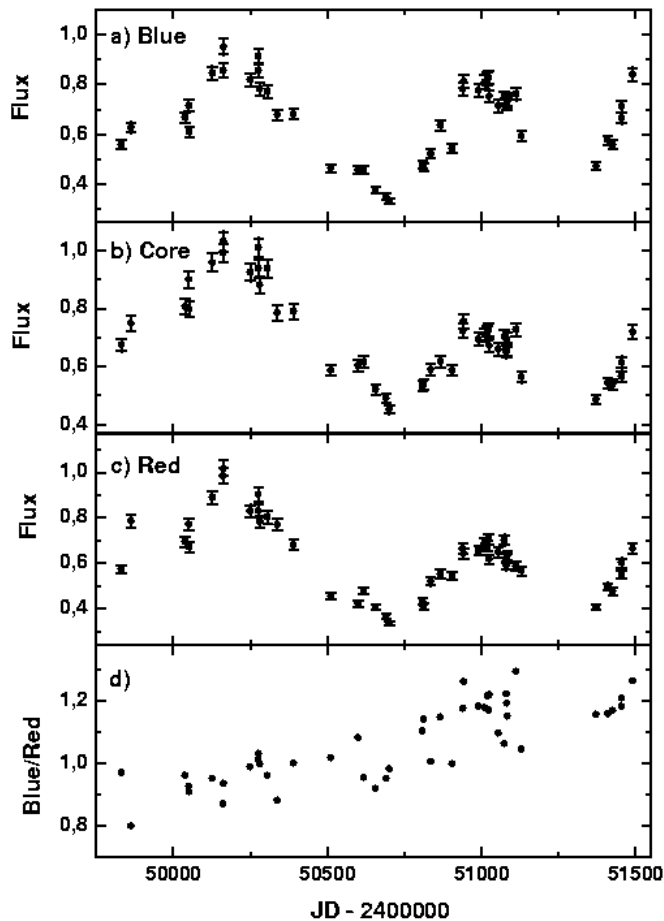}}
   \caption{The light curves. (a)-(c) for the blue wing in the
 5010--5099 \AA\ ($-7100$ km~s$^{-1}$\, to $-1900$ km~s$^{-1}$), interval, for 
the core
 in the 5099--5164 \AA\ ($-1900$ km~s$^{-1}$\, to $+1900$ km~s$^{-1}$), interval,
 and for the red wing of H$\beta$ emission line
 in the 5164--5253 \AA\ ($+1900$ km~s$^{-1}$\, to $+7100$ km~s$^{-1}$), interval,
 respectively. Panel (d) shows the temporal behavior of the the blue-to-red wing flux 
ratio {\it R}}
   \label{fig7}
\end{figure}

\subsection{The $\rm H\beta$ wings}
In order to investigate the flux in the different parts of the H$\beta$
profile, we studied three different velocity bins: a blue
wing ($\rm -7100$\,km~s$^{-1}$\, to $\rm -1900$\,km~s$^{-1}$), a core\, ($-1900$ km~s$^{-1}$\, 
to $+1900$ km~s$^{-1}$) and a red wing ($+1900$ km~s$^{-1}$\, to $+7100$ km~s$^{-1}$). The 
light curves for these are presented in Figures $\ref{fig7}$~a,b,c,
there, we can see that the changes in both the blue and red wings and the core occur 
quasi-simultaneously. This fact implies that, there are predominantly circular 
motions in the region where the broad H$\beta$ emission originates. The blue-to-red 
wing flux ratio {\it R} is also shown in Fig.
$\ref{fig7}$d. Veilleux \& Zheng (\cite{veilleux}) noticed that between 1974 and 
1988, the blue-to-red wing flux ratio followed a fairly smooth, almost
sinusoidal pattern with an apparent period of 10.4 yrs. This
trend was confirmed by Bochkarev {\rm et al.}\,(\cite{boch2}) with observations 
that extended until 1995. Although maxima for the ratio {\it R} were observed in the 
years: 1975, 1985, and 1995, the sinusoidal trend of
{\it R} did not continue during the 1996--1999 period. In Fig. $\ref{fig7}$d, it is 
evident that {\it R} continued to increase in a nearly monotonic fashion between
1995 and 1999.

\subsection {Cross-Correlation Analysis}

\subsubsection {The CCF of the H$\beta$ broad and continuum emissions}

In order to determine a more accurate time lag than the one
described in 3.2, we carried out
a time series analysis using the cross-correlation function (CCF).
 The CCF was calculated by means of the
interpolation method described by Gaskell \& Sparke (\cite{gas2}) and
White \& Peterson (cite{whi}). We
computed both the lags related to the CCF peak ($t_{max}$) and
the CCF centroid ($t_{cen}$). The value of ($t_{cen}$) is that of the center of mass of the CCF for positive delay values. The CCF is calculated by pairing each of the real data points in both time--series with linearly interpolated points for any arbitrary time delay.

\noindent According to Gaskell and Peterson (\cite{gas3}) not every time
series is suitable for the cross-correlation analysis, only those series, for
which the
autocorrelation function (ACF) width is wider
(by at least 10\%) than the width of the corresponding function
for the sampling window (ACFW), contain relevant information about
time lags. The ACFW is computed by repeated
sampling of white noise light curves, convolved with the
observational window. This procedure provides a measure of how much of the
width of the ACF is due to the interpolating scheme, instead of the real
correlation width of a continuous time series sampled at discrete times.
In addition, the half width of the ACF at a zero correlation level determines
the characteristic size of the region from which
variations originate, setting an upper limit to the {\bf BLR} size.

\begin{figure*}
\resizebox{\hsize}{!}{\includegraphics{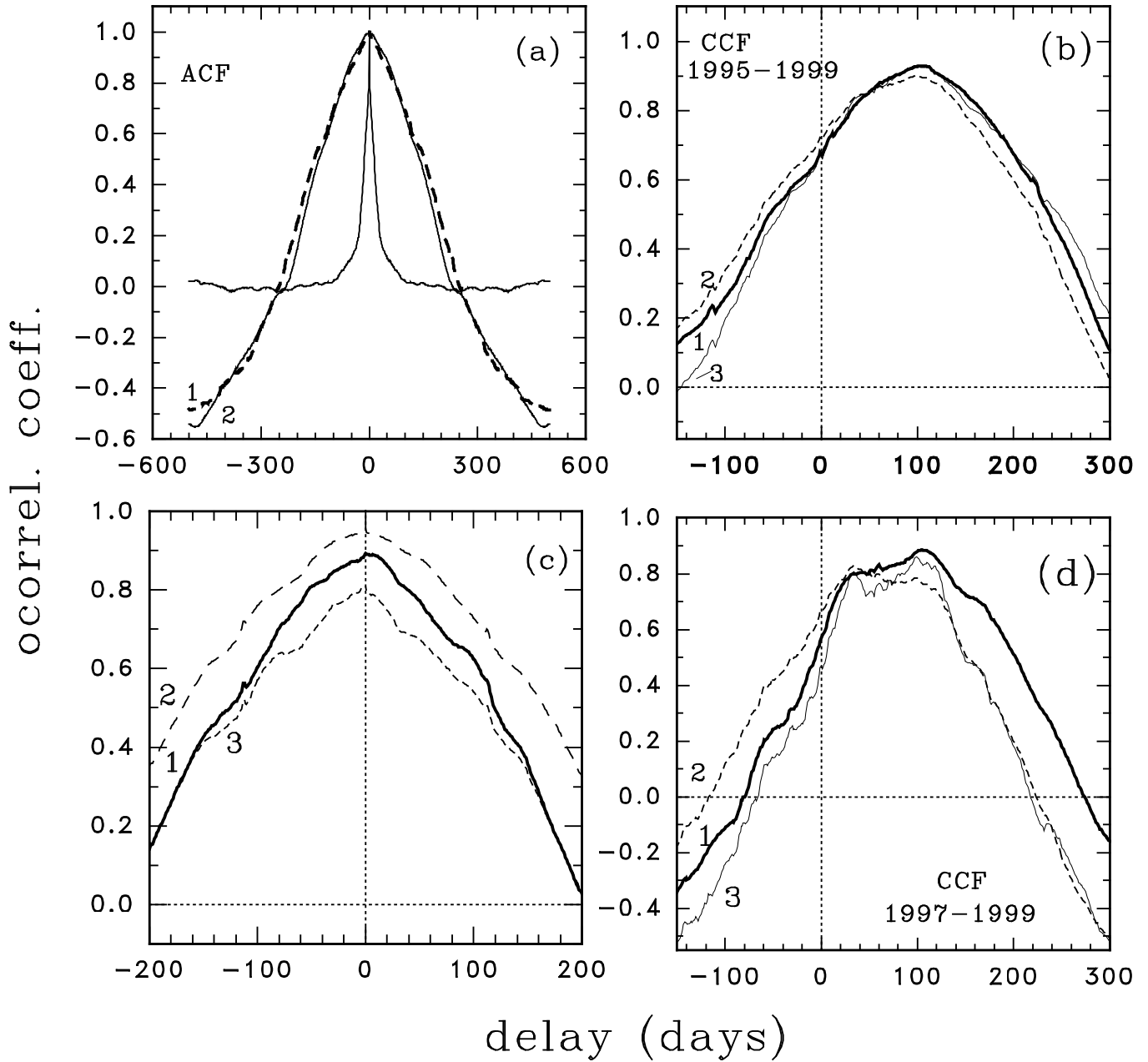}}
\caption{ The Auto- and Cross-Correlation
Functions. In panel {\bf (a)}, the narrow curve represents the
sampling window autocorrelation function (ACFW) for the
spectral continuum and H$\beta$ line. Also, the autocorrelation
functions (ACF) for the H$\beta$ label {\bf (1)} - and the
continuum light curves {\bf (2)} are shown.  On panels
{\bf (b)} and {\bf (d)}, the CCFs for the periods 1995-1999
and 1997-1999 respectively are shown. Curves label {\bf (1)} represent
the CCF between the H$\beta$ flux and the spectral continuum at 5125\AA;
curves labeled {\bf (2)} represent the CCF for the H$\beta$ flux and the
combined continuum and label {\bf (3)} refers to the CCF computed as in
case {\bf (2)}, but for a restricted interpolation of 100
days.  Panel {\bf (c)} shows the CCF between: {\bf (1)} -
H$\beta$ blue and red wings light curves , {\bf (2)} - H$\beta$
red wing and H$\beta$ core fluxes and {\bf (3)} - H$\beta$ blue
wing and H$\beta$ core light curves. In every panel the units
are: delay in days (abscissae) and correlation coefficient (ordinates).} 
\label{ccf} 
\end{figure*}

\noindent The ACFW was calculated for the time series of the
observations, upon which, random Gaussian noise corresponding to 
the observing
errors in the time series was superimposed. The mean ACFW for one 
hundred of
those realizations
is shown in Fig.~$\ref{ccf}${\bf a}. There we can see that the 
ACFW is much narrower than the ACF for either the continuum or 
the H$\beta$ emissions.


\noindent Usually, the time delay is determined either by the location
of the peak, or that of the center of mass of the
cross-correlation function between the continuum and the
emission line light curves. Robinson and P\'erez (\cite{robi})
argued that the position of the CCF peak ($t_{max}$) yields the
time delay corresponding to the inner radius of the {\bf BLR}, while
the position of the center of mass of the CCF ($t_{cen}$)
relates to the size of the {\bf BLR} weighted by luminosity.

\noindent The errors, in the determination of the time lags, 
were estimated
through a Monte Carlo
simulation. From 1000 independent realizations of the CCF, the $t_{max}$
and $t_{cen}$ distributions were obtained. Prior to the
computation, the time-series involved, were corrupted with
Gaussian noise in correspondence to the typical errors in them.
Cutting the distribution functions at a 17\% probability level,
the uncertainty related to a 67\% confidence level (1$\sigma$
error) was determined.

\noindent Given that, during the International Monitoring Programme of the Sy galaxy
NGC 5548, for a period of 8 years, it was
observed by Peterson {\rm et al.} (\cite{pet7}) that the time lag in this
object does not have a constant value, and that changes of the delay
may be related to the luminosity variations of the central source.
And since, during time interval cover by our observations, the photometric and
spectral light curves of 3C 390.3 show two distinct
flares.  It is important to determine the delays not only for
the entire time-series, but also for the segments of it, that
correspond to relevant events in the light curves.
Therefore, analysis of the light curves was carried 
out for various cases, as follows:

The CCF analysis has been carried out separately for the entire data set
-- case A (JD=2449832-2451527), and for the two sub--sets 
(case B, JD=2449832--2450701; and case C JD=2450618--2451527). It should be
mentioned that for case B, we do not have supplementary photometry.
While in case C we have included our broad band photometry. 
For this reason, the flux cross-correlation was calculated
not only for the spectral continuum fluxes
F$_\mathrm{sc}$, but also for the combined continuum data  F$_\mathrm{scv}$. To the latter, we added our photometric data in the V band to the spectral
continuum data as discussed above in 3.2.

Since many spectral observations were
obtained with long time intervals between them, the CCF was
calculated restricting the interpolation processes to time
intervals no longer than 100 days (cases - AD,CD).
That is, prior to the CCF computation, both time series were
analyzed as to whether or not they were suitable for pairing.
With this 100 day restriction in the pairing, possible effects 
associated to long time gaps are reduced.

\begin{figure*}
\resizebox{13.0cm}{!}{\includegraphics{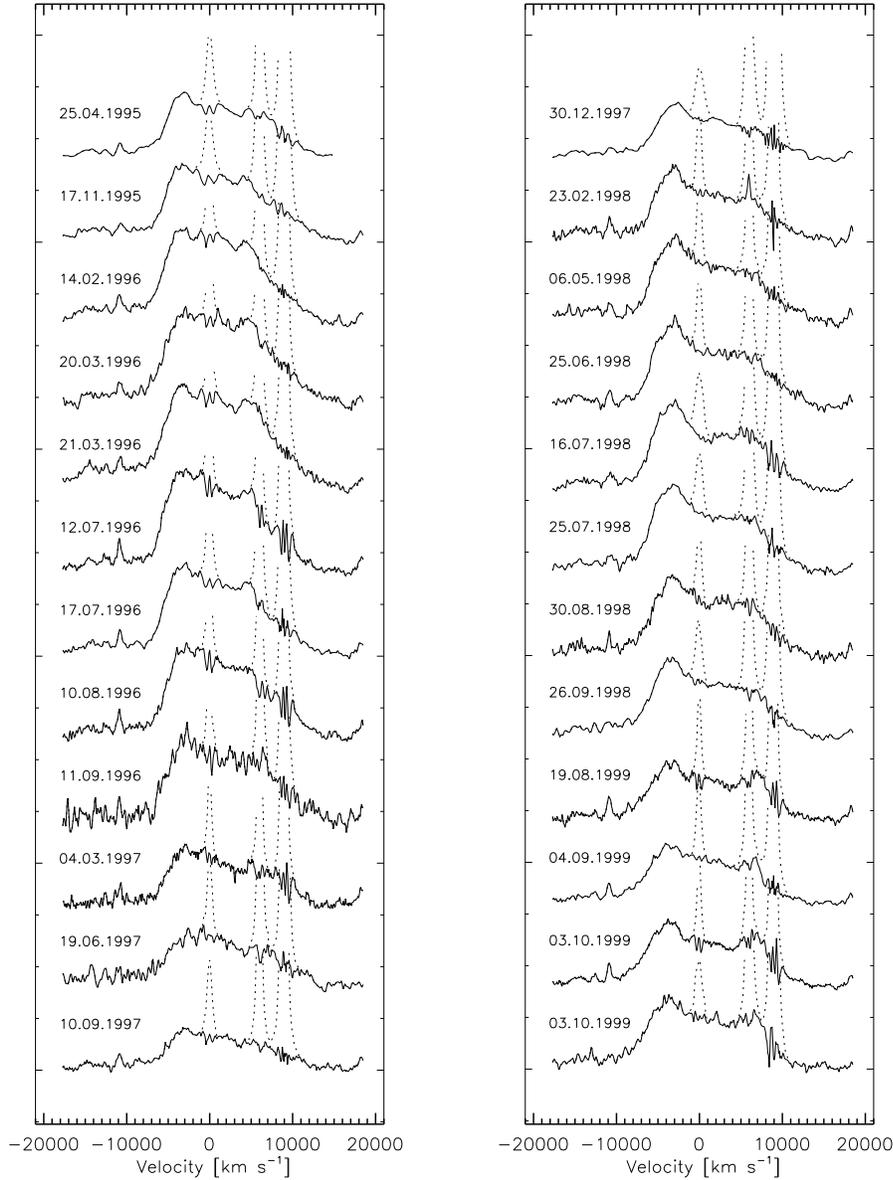}}
\caption{Typical broad H$\beta$ profiles (solid line) after
continuum and narrow components subtraction (dashed line).}
\label{hbprof}
\end{figure*}

The results of the cross-correlation analysis 
are presented in fig. $\ref{ccf}$  and summarized
in Table $\ref{table6}$. There we list: in Col. 1 the time-series case;
 in col. 2 -- the cross-correlation  case:  (F(H$\beta$) -- F$_\mathrm{sc}$) -- between
the continuum flux F(5125) and H$\beta$ flux; (F(H$\beta$) --
F$_\mathrm{scv}$) -- between the combined continuum flux (F(5125)+F(V))
and H$\beta$ flux; in col. 3~--~$t_{max}$ -- the time lag in days,
determined from the position of the CCF maxima; in col. 4 -- 1$\sigma$ interval for
$t_{max}$; in col. 5 -- $r_{max}$-- the cross-correlation coefficient for the CCF maxima;
in col. 6 --$t_{cen}$ -- the time lag in days, determined from the centroids
of the CCF; in col. 7 --  the estimated 1$\sigma$ uncertainty for $t_{cen}$; in cols. 8 and 9 --
the number of points, used in the
computations -- N(H$\beta$)
and --  N$_\mathrm{c}$.

\begin{table*}[hbtp]
\caption{\footnotesize{Time Delays in Days}}
\label{table6}
\hfill
\begin{tabular}{|ll|clc|cl|cc|}
\hline
Time &   Case   & $t_{max}$ & 1$\sigma$& $r_{max}$ &
$t_{cen}$ & 1$\sigma$& N(H$\beta$) & N$_\mathrm{c}$\\
series& & &interv.& & & interv.&& \\
\hline
A & F(H$\beta$)--F$_\mathrm{sc}$ & 105
&& 0.928 & 99 & & 44 &44\\
A&F(H$\beta$)--F$_\mathrm{scv}$&100&97---101&0.900& 86&79---92&44&127\\
AD&F(H$\beta$)--F$_\mathrm{scv}$&99& & 0.930& 98 & & 37& 45 \\
\hline
B&F(H$\beta$)--F$_\mathrm{sc}$& 98 && 0.946 & 109 && 20 & 21 \\
B&F(H$\beta$)--F$_\mathrm{scv}$&99&93---113&0.954&112& 98---114&20&29\\
\hline
C &F(H$\beta$)--F$_\mathrm{sc}$&105 && 0.885 & 90 && 27 & 26\\
&& 34 && 0.800 &&&& \\
C&F(H$\beta$)--F$_\mathrm{scv}$&34&33---37& 0.829& 63&50---72& 27&109\\
&&99&97---119& 0.784 &&&& \\
CD&F(H$\beta$)--F$_\mathrm{scv}$&99& &
0.859 & 78 & & 24 & 33 \\ && 33 && 0.803 &&&&\\
\hline
A&F(H$\beta$)b--F(H$\beta$)c&0&-3.7---1.2&0.809&-1.6&-5.0---1.2 &44&44\\
A&F(H$\beta$)r--F(H$\beta$)c&-1.0&-3.3---0.9&0.957&-1.5&-4.6---0.8&44&44\\
A&F(H$\beta$)b--F(H$\beta$)r&0.5&-1.3---7.8&0.894&0.6&-2.1---8.0&44&44\\
\hline
\end{tabular}
\vspace{0.1cm}
\end{table*}

Inspection of Fig.~$\ref{ccf}$  and 
Table~$\ref{table6}$, shows that:
\begin{enumerate}
\item All cases yield a time lag of H$\beta$
relative the continuum flux changes $t_{max} \approx 100\pm 10$
days or a little less for $t_{cen} \approx 90\pm 10$ days.

\item In the time interval from Jun.1997 - Dec. 1999 (case C),
an additional, shorter delay of 33-37 days is apparent.

\end{enumerate}

Our delay values differ from the results of Dietrich et al.(1998).  
From Fig. $\ref{regr}$ one can
clearly see, that the 20 day lag, obtained by these authors yields
a much larger variance and lower correlation coefficient between line and continuum fluxes than the 100 day lag. In our data we do not see a lag 
of 20 days or something like it. The reason of this discrepancy is 
not clear to us. Perhaps it is related with the longer time interval
covered by our observations. It is conceivable that the time delay in this object does not have a constant value, and that changes of the time lag may be related to the luminosity variations of the central source. A similar case to that of the Sy galaxy NGC 5548 studied by Peterson {\rm et al.}
(\cite{pet7}).
 
\subsubsection{The CCF for different parts of the H$\beta$ profile}
We calculated the cross-correlation of the responses of the blue wing 
($-7100$ to
$-1900$\,km/s), of the core ($-1900$ to $+1900$\, km/s) and of the red
wing ($+1900$ to $+7100$\, km/s) of the H$\beta$ profile to
variations of the optical continuum. The delays found were 100--113 
light days for
$t_{max}$, and 86--102 light days for $t_{cen}$. However, it
should be mentioned that 7\% of the maxima in the distribution of the
CCF peak for the case of the F(H$\beta$) blue wing -- F$_{scv}$
correspond to a delay of about 40 days, almost the same as in
case C previously discussed (see Sect. 3.4.1).

The possible differences in the response of the various parts
of the line profile can be studied through their
cross--correlation functions. In our case, the cross--correlation
of the blue-core, red-core and blue-red H$\beta$ components are 
presented in
Fig. $\ref{ccf} {\bf c}$. The corresponding time lags for the blue 
and red wings
relative to the core, and of the wings relative to each other
are listed in Table $\ref{table6}$.
There, we can see that our CCF analysis for the H$\beta$ wings
does not revealed any delay in the variations of the line wings
with respect to the central part of the line, or relative to
each other. These results preclude the possibility of having
a BLR velocity field dominated by radial motions.

\subsection{H$\beta$ profile changes}

In Fig. $\ref{hbprof}$, we present the profiles of the broad
H$\beta$ component, in velocity units relative to the narrow
component. There the narrow [\ion{O}{iii}] and H$\beta$ emissions
were removed by means of
the Gaussian-fit procedure previously described. Characteristic
features in the profile of the broad H$\beta$ line are the
presence of a blue bump and an extended red wing.
The blue wing of H$\beta$ was
brighter than the red one during the 1995--1999 interval.  In
1996, the intensity of the wings was very
strong. Yet in 1997, the emission in the H$\beta$ wings was rather
weak.  

\begin{figure}
\resizebox{\hsize}{!}{\includegraphics{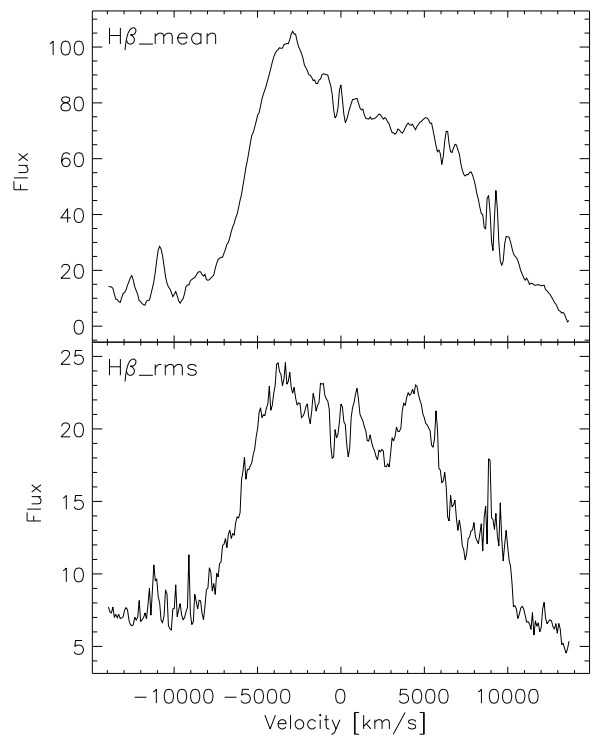}}
\caption{ The mean and rms H$\beta$ profiles in velocity units}
\label{rms}
\end{figure}

\subsubsection{Mean and Root-Mean-Square Spectra}

\noindent The comparison between the average and root-mean-square (rms) 
spectra provides us with a good measure of profile variability. Since, any 
constant contribution to the spectra is excluded from the rms spectrum.
The mean H$\beta$ profile after removing narrow the H$\beta$ and 
[\ion{O}{iii}]
lines  and the absolute rms variations per unit
wavelength are shown in Fig. $\ref{rms}$. Our results are similar 
to those of 
Dietrich {\rm et al.}\, (1998): ie. the mean H$\beta$ profile is clearly 
asymmetrical 
with a full width at zero intensity of about
20\,000 km~s$^{-1}$. The blue bump in the mean spectrum is located 
between 
-3000 and -4000 km~s$^{-1}$\, while the red bump is seen 
between +4000 and +5000 km~s$^{-1}$. The mean profile shows a blue bump 
brighter 
than the red one during the 1995--1999 lapse. From Fig.~$\ref{rms}$ 
(bottom) the FWHM-(rms) is about 12\,000 km~s$^{-1}$.

\subsection{The velocities of the H$\beta$ blue bump}

\begin{figure}[htbp]
\resizebox{\hsize}{!}{\includegraphics{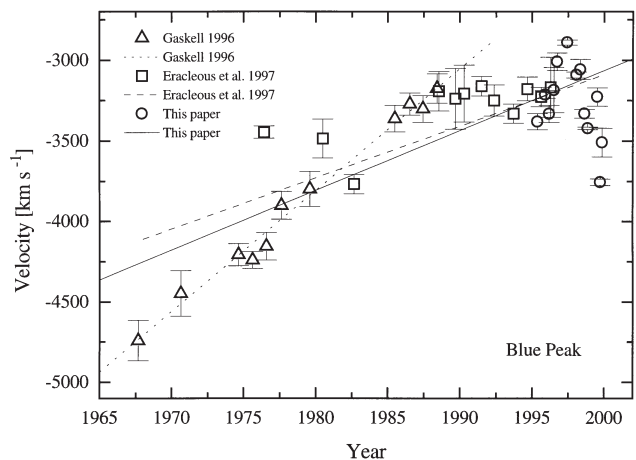}}
\caption{The average radial velocity curve of the blue bump
peak. The different symbols represent: triangles - the annual H$\beta$ average data from Gaskell (\cite{gas4}), open
squares - the H$\alpha$ data from Eracleous {\rm et al.}\, (\cite{erac3}), open circles - our
H$\beta$ data. The different best fit models are explained in the top left
panel.}
\label{fig12}
\end{figure}

We have measured the radial velocities of the emission peak of
the H$\beta$ blue bump relative to the narrow H$\beta$
component from Gaussian fits of the top of the blue 
bump down to the
30\% level of the peak brightness. A method similar to that
applied by Eracleous {\rm et al.}\, (\cite{erac3}). A comparison of
our blue bump radial velocities with theirs,  
for the same epochs show a very good agreement between both sets of 
observations. The mean difference being 
about $\pm$60km~s$^{-1}$, a value close to the uncertainties 
of the procedures adopted. Our results are presented, in graphic form 
(open circles),
in Figure $\ref{fig12}$. Each point represents an average of 
several spectra obtained during a lapse of a few months. The 
absolute value of the radial velocity
of the blue bump presents a minimum, observed in 1997, at
$\approx $ $-$2900 km~s$^{-1}$, and a maximum occuring in October 1999 at
$\approx $ $-$3700 km~s$^{-1}$ . The velocity shift of the blue bump
reached a value of about 800 km~s$^{-1}$ during this time.
The changes of the blue bump velocity between 1967 and 1999, taken
from the literature, are also presented in Fig. $\ref{fig12}$.
The short dashed line represents the best fit of the double-line
spectroscopic binary model by Gaskell (\cite{gas4}), while, the long
dashed line represents the fit by Eracleous {\rm et al.}\, (\cite{erac3}). 
Onecan see that our data are reasonably well described by the
latter. These authors have shown that, if the displacement of
the blue bump peak velocity originated from individual broad-line
regions, associated with a massive binary BH, then the infered
rotation period would be about 800 years, and the associated
mass would then be larger than 10$^{11}$M$_{\sun}$. This value for
the BH mass is much larger than the expected by comparison with
other BH candidates found at the center of galaxies.  Therefore,
these authors rejected the binary BH interpretation.
Our results further support their conclusions.

\subsection{The $\rm H \beta$ difference profile}

\begin{figure}
\resizebox{\hsize}{!}{\includegraphics{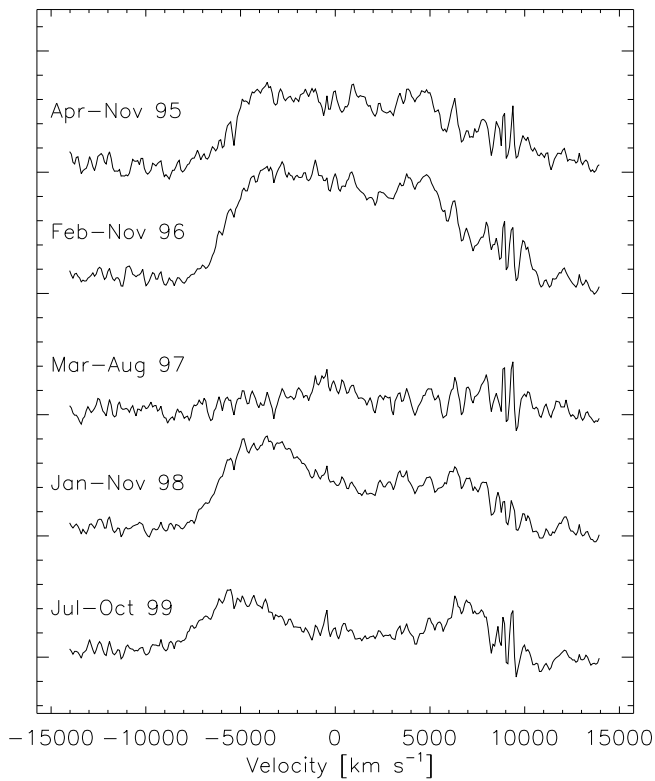}}
\caption{Annual averages from 1995 to 1999 of the difference profiles for the broad
H$\beta$ line.}
\label{fig14}
\end{figure}

We obtained th H$\beta$  difference profiles
by subtraction of the minimum-activity average spectrum
(9 September, 1997) from individual spectra. This
is done after continuum, narrow H$\beta$ and
[\ion{O}{iii}] emission had been substracted as well.
When the spectra for subsequent nights, presented small differences
ie. H$\beta$ integral flux remained constant within 3 to 5\%, they 
were averaged and the spectra of low signal to noise ratio (S/N$<20$) were
excluded from the analysis. The annual average differential
profiles of the broad H$\beta$ line are shown in fig. $\ref{fig14}$.
In order to determine the blue and red peak
velocities, the mean difference profiles were fitted with three
Gaussian functions; one to the blue bump, one to the core, and one
to the red bump.
The derived differences, of the peak velocity for the red and blue
bumps, V$_\mathrm{r}$-V$_\mathrm{b}$, are plotted in Fig.~$\ref{hb-c}$ {\bf c}.
Despite the errors in the determination of the
velocities, a distinct anticorrelation is observed
between  the difference V$_\mathrm{r}$ - V$_\mathrm{b}$
and the flux changes in both H$\beta$ and continuum emission, as it is 
clearly seen in Fig.~$\ref{hb-c}$.

\begin{table*}[hbtp]
\caption{Radial velocities of the  blue bump, red bump and core of H$\beta$ derived from the difference profiles, jointly with contemporary H$\beta$ fluxes.}
\footnotesize
\label{meanvel}
\begin{tabular}{lrrrr} \hline \hline
Velocity (km~s$^{-1}$) &&                             Time Interval\\
Fluxes  &   Apr.1995-Nov.1996 & (Mar.-Aug.)1997& (Jan.-Oct)1998& (Jul.-Oct.)1999
\\
\hline
V$_\mathrm{blue}$ &     -3204$\pm$187     &    --          &   -3867$\pm$75   & -5215$\pm$118
\\
V$_\mathrm{red}$  &      4899$\pm$155     &    --          &    5361$\pm$134   &  7001$\pm$106
\\
V$_\mathrm{core}$  &       617$\pm$187     &   76$\pm$249       &     335$\pm$235   &  1170$\pm$244
\\
V$_\mathrm{r}$-V$_\mathrm{b}$ &     8103$\pm$243      &    --          &    9228$\pm$154   & 12216$\pm$158
\\
$F(\rm H\beta)^{\rm \star}$ &  2.804$\pm$0.37     & 1.733$\pm$0.14    & 2.367$\pm$0.21   & 1.932$\pm$0.25\\
$F_{con}^{\rm \star\star}$& 2.132$\pm$0.54     & 1.035$\pm$0.13    & 1.611$\pm$0.14   & 1.406$\pm$0.17
\\
\hline
\end{tabular}
\begin{list}{}{}
\item[$^{\rm \star}$] -- in units of 10$^{-13}$ ergs s$^{-1}$cm$^{-2}$
\item[$^{\rm \star\star}$] -- in units of 10$^{-15}$ ergs s$^{-1}$cm$^{-2}$A$^{-1}$
\end{list}
\end{table*}

The annual mean velocity of the blue, red and core
components derived from the Gaussian analysis, jointly with
the V$_\mathrm{r}$-V$_\mathrm{b}$ velocity difference of
the bumps, are listed in Table~$\ref{meanvel}$. There we notice that, the radial velocity of the blue component increased in absolute value
from $-3200$ km~s$^{-1}$ in 1995-1996 to $-5200$ km~s$^{-1}$\, in 1999.
At the same time the radial velocity of the
red component increased from $+4900$ km~s$^{-1}$ in 1995-1996 to $+7000$ km~s$^{-1}$\, 
in 1999. Then, the absolute velocity of both components
increased by about 2000 km~s$^{-1}$ during this period of time.  
That is, the difference V$_\mathrm{r}$-V$_\mathrm{b}$
increased by about 4000 km~s$^{-1}$.
Table~$\ref{meanvel}$ lists the observed mean fluxes of H$\beta$ (F(H$\beta$))
and continuum ($F_{con}$) light in the same time intervals. It is evident
that the lower annual average velocities of the blue and red components,
or their difference correspond to larger mean flux values.
In the frame of an accretion disk model, the optical continuum changes
are a consequence of changes of the luminosity of the central X-ray source,
causing a variable disk irradiation.

The observed velocity variations could be explained if, at different times,
the zone that
contributes with the maximum luminosity to H$\beta$, changes in radius
within the disk. The position of this zone will depend on the luminosity
of the central source. Then, it is likely that during 1995--96, we were observing 
enhanced
H$\beta$ emission from parts of the disk which are located further away from
the BH ie, when the luminosity of the central source is higher, and consequently heated the outer regions of the disk. This
results 
in a maximum emission zone located at larger disk radii.
Later on, in 1999, maximum emission came
from parts of the disk located at a smaller radii, when the luminosity
of the central source was much lower.
In the case of a circular disk the ratio of the maximum H$\beta$ emission
radii must be proportional to the inverse squared velocity.
Hence, if we consider the velocity values listed in Table~$\ref{meanvel}$
for 1995--96 and
the average velocity values for 1998--99, the ratio of the radii of maximum
H$\beta$ emission is about $1.77 \pm 0.2$. This is in
excellent agreement with the ratio of the size of the emitting zones
as inferred from the lag time ratio of the centroids of the CCF (ie. $1.78 \pm 0.3$, see Table~$\ref{table6}$).

These transient phenomena are expected to result
from the variable accretion rate close to the central black hole.
Note that from the theory of thin accretion disks
(Shakura \& Sunyaev \cite{shak}) one can calculate the energy dissipated through
viscosity, the resulting spectrum emitted by the disk, and time
scales in which, the emission from the disk varies in response to
a variable accretion rate. These are rather long ($\approx$  $ 10^5$  years
for a $ 10^7\, \mathrm{M_{\sun}}$ BH ). However, as the accreted matter approaches
the BH, at a distance of few tens of gravitational radii,
a large fraction of the accretion energy is transformed into X-ray radiation.
The rapid variability of this radiation on time scales of hours, days and weeks is
due to changes in the final accretion rates.

Currently, a model in which the surface temperature of the disk is
modulated by the irradiation caused by a variable luminosity of the X-ray source
has been
suggested by several authors (see Ulrich, \cite{ulr} and references therein).
In some models it is assumed that the rapid variability (days, weeks) of
the continuum X-ray source can be explained by the fact that
a significant fraction of
the accretion energy near the black hole is spent on heating a corona
by a mechanism reminiscent of flares in the solar corona (Galeev et al. \cite{gal}).
Hot electrons, in turn, transfer some
of their energy to ambient soft X-ray and UV photons (some emitted
by the disk), producing medium energy and hard X-ray radiation via
inverse Compton emission.

\subsection{Modeling of the H$\beta$ Broad Components}

As mentioned above, several models have been
proposed to explain the double-peaked line profiles. Among them, the
emission by an accretion disk or a torus, whose presence seems
essential to fuel a black hole. Our study favours the formation of the broad H$\beta$ 
line of 3C 390.3 in an accretion disk. Superluminal motions have been observed at radio 
frequencies (Alef et al. \cite{Alef}), these are indicative
of the presence of a relativistic jet, with its
axis aligned close to the line of sight (Orr \& Browne, \cite{orr}).
Therefore, the disk's axis is probably close to the line of sight.

Under the assumption of an accretion disk as the site of emission line
formation, the study of the profile changes with time, may provide relevant information, even 
if we do not advance any hypothesis about the disk structure or to how it is
illuminated. Moreover, we could
reject the disk hypothesis, if we find some inconsistency in the results (for
instance a change of the BH mass or the disk inclination).

We assumed that: i) the Balmer lines are emitted by a disk rotating about
a central object with Keplerian velocity, and
ii) the disk emissivity varies with the distance $r=R/R_g$ to the 
center, where $R_g$ is the
gravitational radius ($2GM/c^2$) with $M$ being the mass of the BH.
Then we computed the profiles emitted by such disk, taking into account the
relativistic Doppler and gravitational redshifts, using the equations given by
Gerbal \& Pelat (\cite{gerbal}) and Chen et al. (\cite{chen}). These equations are valid for 
a disk seen at
inclination angles $i$ smaller than $80 \deg$, which is the case for 3C390.3.
The resulting profiles are double-peaked and asymmetric. They were
convolved with the instrumental profiles and fitted
to the observed ones, using the Davidon-Fletcher-Powell method
found in Minuit's package (James \cite{james}).

One important parameter in the fitting procedure is the inclination angle $i$, which must
remain constant. For a simple power-law for the disk emissivity
ie. $F(\rm H\beta)\propto r^{-q}$, the line wings cannot be
properly reproduced, a fact already noticed by Chen et al. (\cite{chen}) for Arp 102. The
profile fits are improved significantly, when a double power-law
emissivity function is adopted. In this case, the important fitting parameters of the model are: the
indices $q_1$ and $q_2$, the radius $r_1$ at which the slope changes from
$q_1$ to $q_2$, and the outer radius of the disk, $r_{out}$ (all radii being
expressed in units of $R_g$). The inner disk radius does not play an important role. 
The four relevant parameters are expected to vary with the continuum flux level
and the illumination of the disk.

In this paper, we present only a few of the results, a more detailed description
of the model and the results for the H$\alpha$ and H$\beta$ profiles will be
presented in a separate paper.

\begin{figure}
 \resizebox{\hsize}{!}{\includegraphics{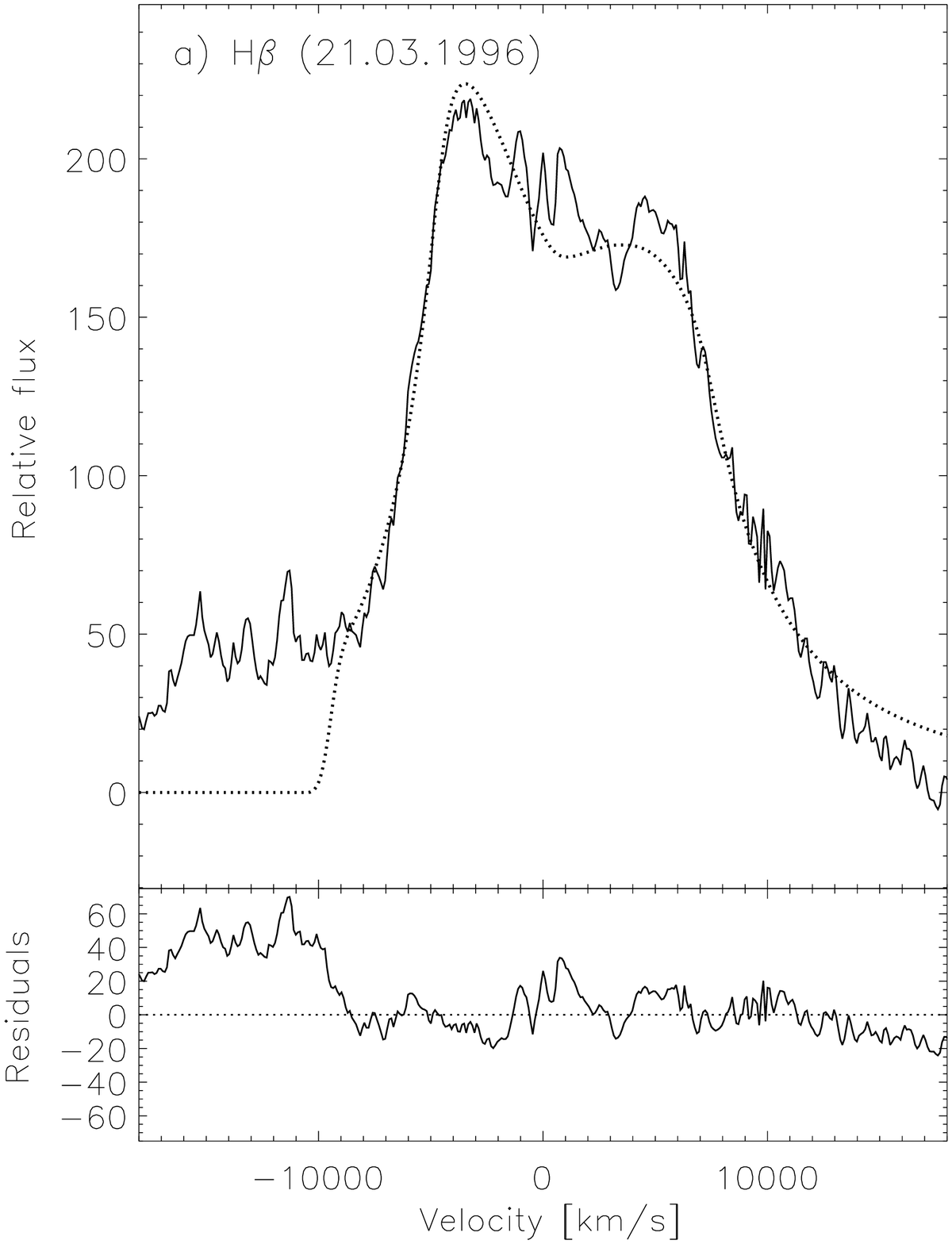}\includegraphics{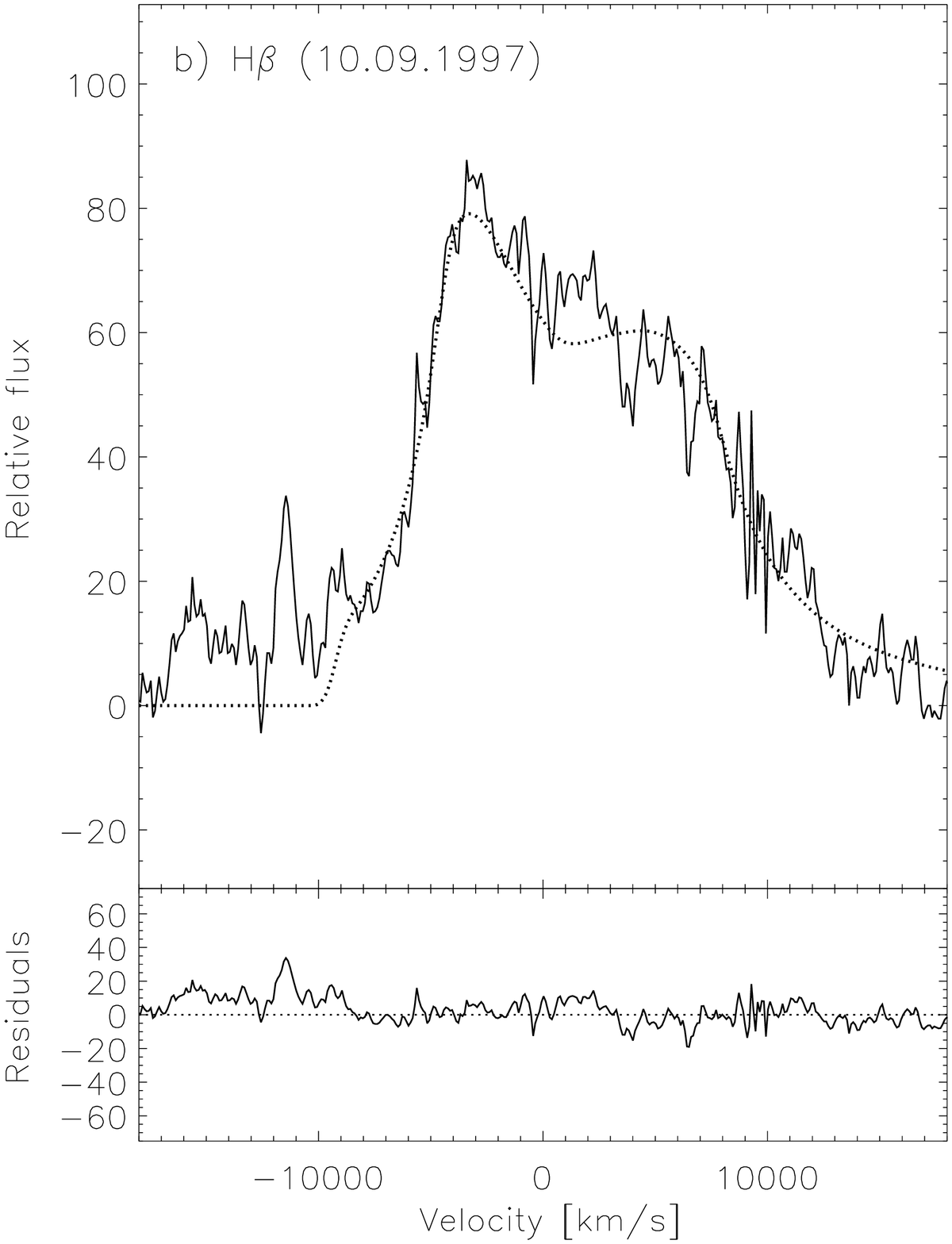}}
 \resizebox{\hsize}{!}{\includegraphics{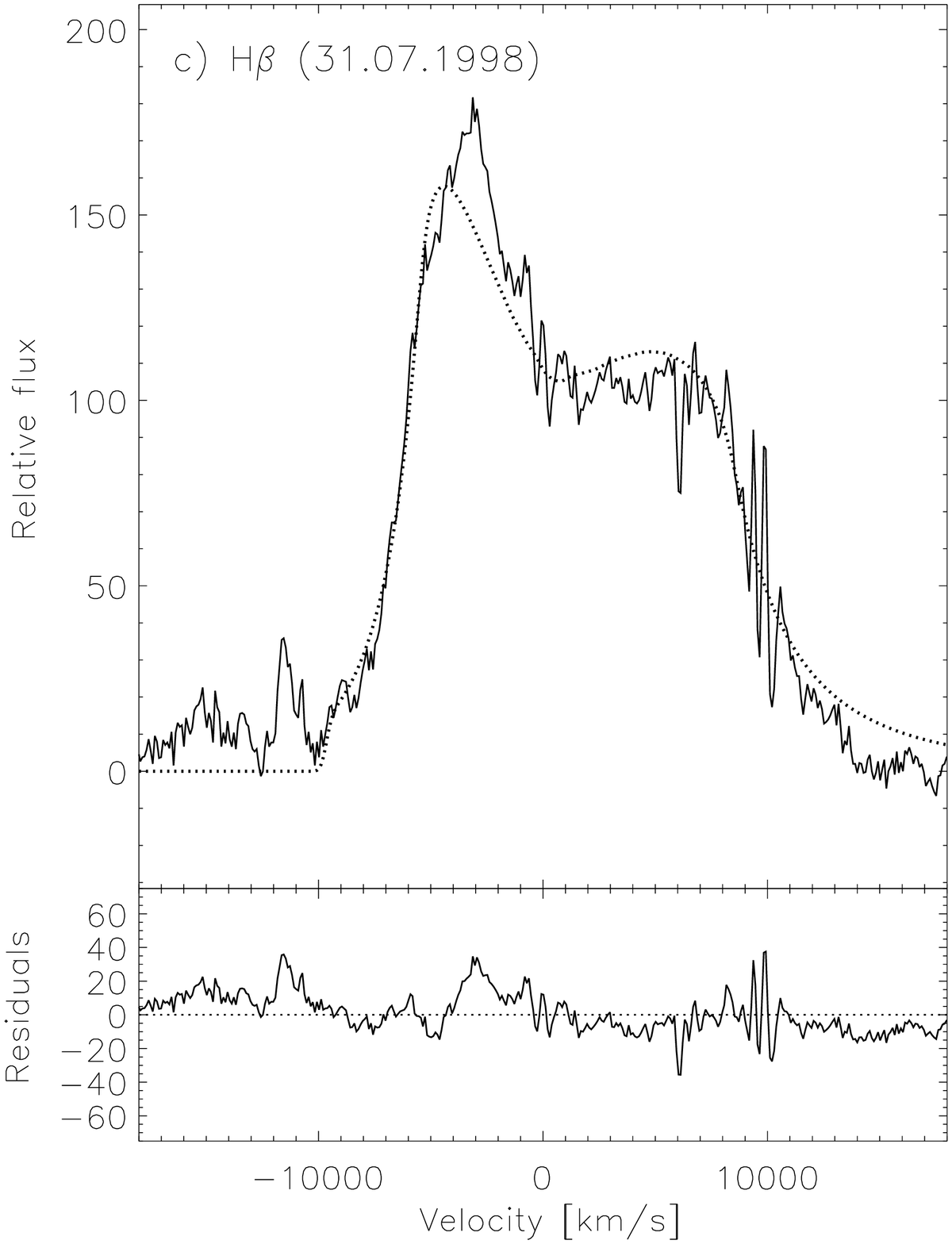}\includegraphics{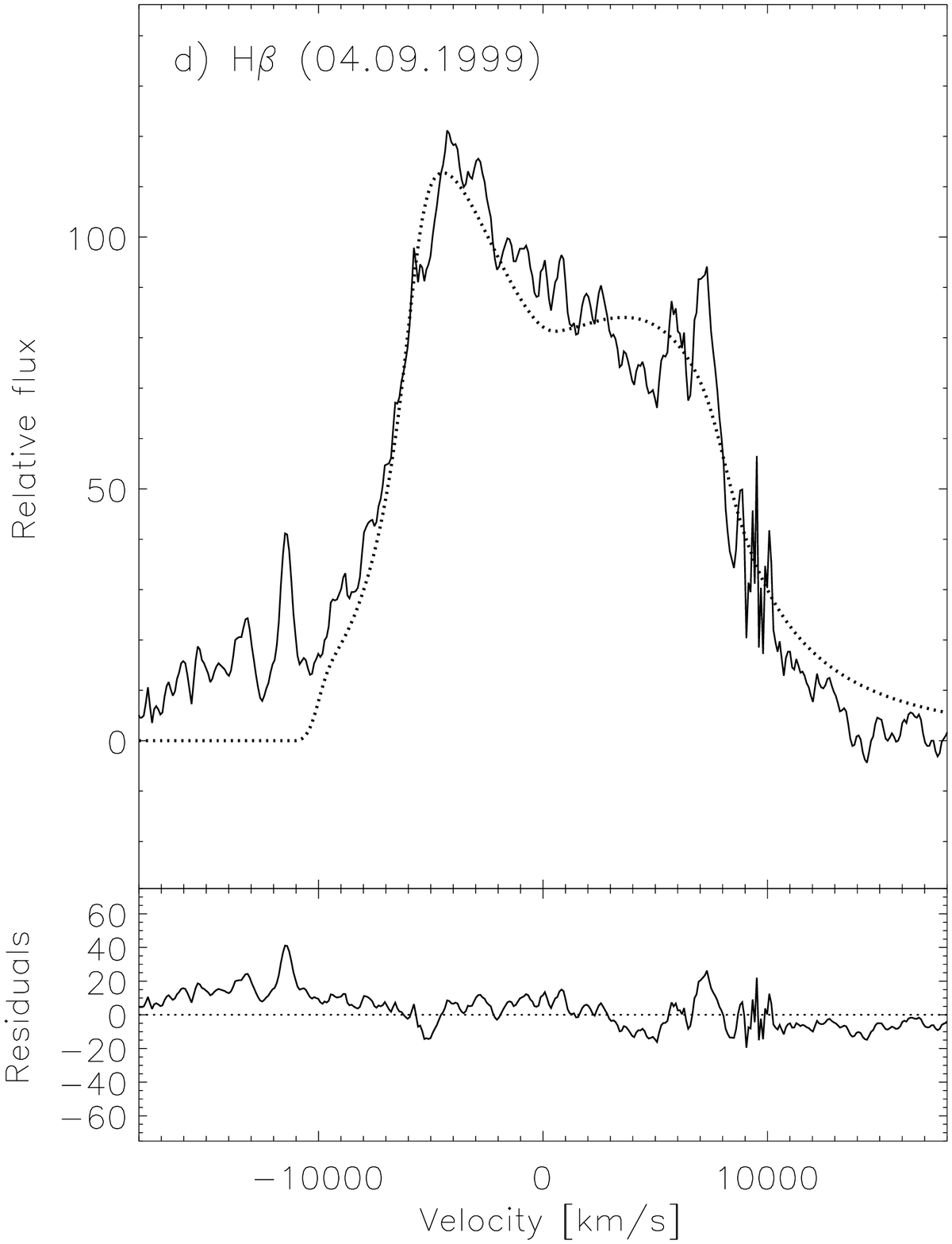}}
\caption{The fits to the H$\beta$ observed profiles for different
epochs, by an accretion disk model with an inclination angle of about 
25$^{\circ}$,
and with a region of maximum emission at about 200 gravitation radius. The
dotted line represents the model profile, while the full thin line the observed
one. In each window, the lower panel represents the residuals for the following
dates: JD=24410163, 24410701, 24411025, 24411426.}
 \label{mod}
\end{figure}

The fits of four selected H$\beta$ profiles, distributed in time, and
representative of maximum and minimum activity states, are shown in Fig.~$\ref{mod}$.
From our fitting procedure we find that: i) $i=25 \pm  3 $ degrees,
consistent with a constant value of $i$, and ii) surprisingly,  despite the large 
changes in the continuum and line fluxes, $q_1\approx 1$ and $r_1\approx 200R_g$ do not 
vary significantly, while $q_2$ varies between 3.46 and
3.51 for the fits to the four profiles.

One should seek a physical explanation of this behavior. According to the results, the bulk of the line
emission is produced at a radius close to $r_1$ i.e. at $R\approx 200R_g$.

It is interesting to note that this value corresponds to the
radius where the accretion disk becomes gravitationally unstable (cf.
Collin \& Hur\'e \cite{col2}). At larger radii, the emissivity
decreases very rapidly, faster than it would be expected
in the case a continuum source illuminating the accretion disk, or a system
of rotating clouds having a constant density with a canonical BLR value of
$n \ge 10^{9}$ cm$^{-3}$. In this case, $F$(H$\beta)$ would
roughly be $\propto F_{inc}^{0.5}$ (cf. for instance
Dumont, et al. \cite{dum3}), where $F_{inc}$ is the flux incident on the disk.
And since at large distances
from the continuum source $F_{inc}\propto r^{-3}$, one would find
that $F$(H$\beta)$ $\approx r^{-1.5}$ for a constant density. If the disk is
geometrically thick (a ``torus''), the exponent is smaller, $F_{inc}$
being proportional roughly to $r^{-2}$. This means that the density
decreases rapidly with increasing radius, as about $r^{-2}$, as expected
if the disk is in a state of marginal instability (Collin \& Hur\'e, \cite{col1}).
On the other hand, for smaller radii, the emissivity varies slowly with
$r$, which is compatible with the disk irradiated by the central source
of continuum if the density is larger than $10^{9}$ cm$^{-3}$.
Thus this simple change of the density in the accretion disk could induce a
``physical radius'' of the emission region quite independent of the
value of the continuum flux, and could explain why the radius of
maximum emission does not vary strongly during the monitoring.

So, our study favours the formation of the broad H$\beta$ line of 3C 390.3
in an accretion disk.

\subsection{The mass of the  Black Hole in 3C~390.3}

Using the virial theorem jointly with the reverberation-rms method
(cf. Peterson et al. \cite{pet6b}), the mass of the black hole in 3C390.3
can be estimated from the relationship of Wandel et al., (\cite{wand}):

$$  M_{rev}\approx (1.45\times 10^5 M_{\sun})
\left(\frac{c \tau}{\rm lt-day}\right)v^2_{rms,3},  (4) $$
\noindent
  where $v_{rms,3}=v_{FWHM}({\rm rms})/10^3$ km\,s$^{-1}$.

Using our results: $v_{FWHM}({\rm rms})\approx 12\,000$\,km\,s$^{-1}$ (Fig.~$\ref{rms}$b),
$(c\tau/{\rm lt-day})\approx 100$ (Table~$\ref{ccf}$), we obtained
$ M_{rev}\approx 2.1\times 10^9\,M_{\sun}$. This value is substantially larger 
than
$M_{rev}\approx 4\times 10^8\,M_{\sun}$, obtained by Wandel et al. (\cite{wand}). The difference is due to the fact
that the determination of Wandel et al. is based upon: 1) a time lag of 24 days,
instead of the 100--day lag found in the present study; 2) a the value for
$v_{FHWHM}({\rm rms})\approx 10500$\,km\,s$^{-1}$ instead of our value for
$v_{FWHM}({\rm rms})\approx 12\,000$\,km\,s$^{-1}$. The upper limit for the mass 
of the BH in 3C 390.3, estimated from X-ray
variability is $M(BH)< 1.5\times 10^{10}\,M_{\sun}$ (Eracleous et al.
\cite{erac3}).

Assuming that  $M_{rev}\approx M(BH)$, we estimate a value for
$R_g$ (the BH gravitational radius),
for the derived value of the BH mass:

   $R_g= 2GM(BH)/c^2 \approx 6.2\times 10^{14}$\,cm.

According to our model calculations (Sect. 3.8 ), the bulk of H$\beta$
emission is produced at $r\approx 200\,R_g$, which corresponds to a time lag
$\tau= r/c\approx 48$\, light days.
This implies a shorter time lag than the one derived in this paper (Table~$\ref{ccf}$).
Can these values be reconciled?
The reverberation method assumes that the FWHM of the lines reflects virialized motions,
while we assumed that the motions are purely rotational.
These two working hypotheses may explain differences
in the size of the emitting region on the order of $\approx 40-50$\,\%. Then the time lag for 200\,$R_g$ could be about $\tau\approx 70$\,lt.days. A value which is closer to 
the mean lags listed in Table~$\ref{ccf}$: $\tau_{cen}\approx (86\pm 24)$; and $\tau_{max}\approx (67\pm32)$. Worth, mentioning is the good
agreement (within 20\ 
"cool dense material" reprocesses X-ray radiation into the Fe Kalpha line
detected by Eracleous, {\rm et al.}\, (\cite{erac2}).

\section{Conclusions}

The results of a 4-year (1995--1999) spectroscopic and
broad-band BVRI photometric monitoring of the AGN 3C 390.3 are
presented in this paper for the H$\beta$ spectral region.

Our main conclusions are:
\begin{enumerate}
\item The historical light curve in B shows a large
increase of brightness during 1970--1971 followed by a gradual decrease
until 1982. At that time, the nucleus of 3C~390.3 was in a minimum
luminosity state. On the descending branch of the light curve,
flares with an amplitude up to $1^m$ were observed.

\item The broad H$\beta$ and continuum fluxes vary by a factor
of about three during the time interval 1995--1999.

\item Two large amplitude outbursts in the intensity of the H$\beta$ line
and the continuum, 
with different duration were observed: (Fig.~$\ref{hb-c}$): in (Oct.1994-Jul.1997) - a 
brighter flare lasting $\approx $1000 days and another one in  
(Jul. 1997 -- Jun. 1999) with that lasted $\approx $700 days. The event duration is 
defined as the interval between minima in the light curve.

\item From a cross-correlation analysis, a delay of $\approx $ 100 days
in the response of the H$\beta$ emission with respect to the 
optical continuum changes is obtained from the observations at
any given time. Yet, for observations obtained between the June 1997
until 1999 (JD2450618 - 2451527 -- case C), a period in which the sampling
was better, a shorter lag of 33 to 36 days appears jointly with the one of $\approx $\,100 days. In the light of
these results, we have to admit that either, our data set is not
suitable for an accurate determination of the delays due to a
somewhat poor temporal sampling, or that the delay for 3C 390.3
varies with time, i.e. the correlation between the flux
variations in the continuum and in H$\beta$ at different times
is not the same. In the 8 year monitoring campaign of NGC 5548,
Peterson et al. (\cite{pet7}) have determined, that the time delay
had varied. They suggested that changes in the time
delays may be related to luminosity variations of the central source.
Yet, in other investigation of NGC 5548 (Netzer and Maoz, \cite{netz}) have
shown, that continuum variations with a long time scale --long lasting flares-- yield longer time lags. Both posibilities are present in our
results for 3C 390.3 ie.: a) the continuum outburst
of 1994-1997
with larger amplitude and duration ($\approx$ 1000 days) yields a longer
time lag (100 days); b) the continuum outburst in 1997-1999 of a smaller
amplitude and duration ($\approx$ 700 days) yields two lags: 100 and 35 days.
We believe that both, the optical continuum and $\rm H\beta$ flux variations 
reflect
the changes in the X-ray irradiation, modulated by accretion rate changes.
And that the radius $\rm H\beta$ emission zone, changes due
surface temperature changes in the disk (Ulrich, \cite{ulr}).

\item The observed flux for the blue and red wings of the 
profile varied quasi-simultaneously (Fig. $\ref{fig7}$) 
(i.e. whenever the flux
in the blue wing increases, that in the red wing also
increases). We have not detected a delay in the changes
of the H$\beta$ blue wing ($-$7100\,km~s$^{-1}$ to $-$1900\,km~s$^{-1}$)
and of the red wing
(+1900\,km~s$^{-1}$ to +7100\,km~s$^{-1}$) with respect to the core of the 
line
($-$1900\,km~s$^{-1}$ to +1900\,km~s$^{-1}$) or relative to each other. 
A similar
result was obtained by Dietrich et al. (\cite{die2}) from monitoring
data for 3C 390.3 during 1994--1995. These results exclude the
presence of considerable radial motions in the BLR in
this object.

\item The blue-to-red wing flux ratio shows a monotonic
increase with time for the period covered by our observations
(Fig. $\ref{fig7}$d). The quasi-sinusoidal changes
(period $\approx $ 10 years) of the blue-to-red bump flux ratio
detected by Veilleux {\rm et al.}\, (\cite{veilleux}) are not present in our 
data for the  1995--1999 lapse.

\item The shape of the H$\beta$ profile has changed
drastically: during 1996, the blue and red bump emission was very strong,
while when the source was faintest (1997) the profiles had rather 
flat tops (Fig. $\ref{hbprof}$). The blue wing was brighter than the red one 
during the observed period.

\item The changes of the radial velocity of the blue bump
relative to the H$\beta$ narrow emission in 1995 until
Oct.~1999 follow the same trend found by Eracleous {\rm et al.}\, (\cite{erac3}).
Their model for a binary BH with a very large mass value
($>10^{11}$M$_{\sun}$) seems to fit our data as well. 
These authors argued that such a large value for the binary BH
mass, is difficult to reconcile with other observations
and theory. Therefore, they reject the hypothesis of
a binary black hole for 3C 390.3. Our results provide further
support to the dismissal of the binary BH hypothesis for 3C 390.3,
on the basis of the masses required.

\item The difference profiles of the H$\beta$ broad
emission revealed that the radial velocities of the blue and
red bumps in the profiles varied, showing an increase in
absolute value by about 2000 km~s$^{-1}$ during the observed period. 
From these results (Table~$\ref{meanvel}$, Fig.~$\ref{fig14}$),
it is infered that an annular zone in the accretion disk with enhanced H$\beta$
emission could have moved towards smaller
radii between 1995 and 1999, i.e. larger Keplerian velocities as the flux in the continuum decreased (Table~$\ref{meanvel}$).
These results are in agreement with a model in which the surface temperature
of the disk is modulated by the irradiation of a variable X-ray source (see Ulrich, \cite{ulr} and references threin).
The X-ray variability on time scales of hours, days,
weeks is, in turn, modulated by accretion rate changes.
Thus, the peak velocity shifts as different profiles 
are also expected to result from the variable accretion rate 
near the black hole.

\item We have fitted the broad H$\beta$ line profiles in the
framework of an accretion disk model. We obtained a satisfactory
agreement for a disk with an inclination angle of about 25$^{\circ}$
and with the region of maximum emission located at about 200
$R_g$. The radius of this zone is comparable to the one
in which the Fe Kalpha line emission originates according to Eracleous {\rm et al.}\,(\cite{erac2}).

\item  The mass of the black hole in 3C390.3 estimated by the
combined
reverberation-rms method (Wandel, et al., \cite{wand})
is $M_{rev}\approx 2.1\times 10^9\,M_{\sun}$.

\end{enumerate}

Our results do not support either the models of outflowing
biconical gas streams or those of
supermassive binary BHs. Instead, we conclude
that our results favor the accretion disk model.

The velocities of the blue and red bumps and their difference, obtained
from $\rm H\beta$ difference profiles, 
show a distinct anticorrelation with continuum flux changes. 
Taking into consideration the time lag and the $\rm H\beta$ flux,
we find that 
the zone of maximal contribution to line emission, moves across the face of 
the disk.
It is located at smaller radii when the flux in the continuum decreases
(bump velocities increase), and to larger radii when the continuum flux
increases (bump velocities decrease).
These transient phenomena are expected to result
from the variable rate of accretion close to the black hole.
At low accretion rates, the flux from the central source may
decrease by a large factor, and the integral flux in the line could decrease.
Then, the maximum emission zone of the broad Balmer lines could shrink to such 
small radii, that the lines could become extremely broad and of low-contrast.
Consequently, at those times, they would be undetactable. This may be the
reason why the broad emission lines were not 
seen in the 3C 390.3 spectrum in 1980 (Heckman {\rm et al.}\, \cite{heckman}) 
or were 
very weak in 1984 (Penston \& Perez\,\cite{pen2}). At those times the 
central source brightness was at long lasting minima (Fig.~$\ref{slc}$).

It is intresting to note, that the model calculations of Nicastro (\cite{nic})
point to a relationship between FWHM broad emission
lines and accretion rate for a given BH mass ---
the lower the accretion rate the greater the line width.
For different BH masses, there is a minimum value of the accretion
rate, below which no broad lines are formed. The permitted interval of 
velocities
ranges from $\approx 20\,000$\,km~s$^{-1}$ for sub-Eddington accretion rates 
to $\approx 1000$\,km~s$^{-1}$ for Eddington accretion rates (i.e. the 
lines with $\rm FWHM>20\,000$\,km~s$^{-1}$ do not exist for any mass of 
the black hole).

\begin{acknowledgements}

The authors want to thank M. Eracleous for providing
some of the data used in Fig.10. We would like to thank
Didier Pelat for his help in the computation of the model profiles.
We are thankful to G.M. Beskin for many useful discussions,
to S.G. Sergeev for allowing us to use some of his
software for spectral analysis tools,
to O. Martinez for some spectral observations and to V.E.Zhdanova for help
in data processing.
This paper has had financial support from
INTAS (grant N96-0328), RFBR (grant N97-02-17625, grant N00-02-16272a),
scientific technical programme ``Astronomy'' (Russia), RFBR+CHINE
(grant 99-02-39120) and CONACYT research grants G28586-E, 28499-E and 32106-E
(M\'exico).\\

\end{acknowledgements}


\begin{thebibliography}{}

\bibitem[1988]{alef} Alef, W., Gotz, M.M.A., Preuss, E., et al., 1988, A\&A 192, 53
\bibitem[1996]{Alef} Alef, W., Wu, S.Y., Preuss, E., et al., 1996, A\&A 308, 376
\bibitem[1995]{alloin} Alloin, D., et al., 1995, A\&A 293, 293
\bibitem[2000]{Amirkhanian} Amirkhanian V. et al., 2000, Bull. Spec. Astrophys. Obs. 50, (in press)
\bibitem[1983]{antokhin} Antokhin, I. I.; Bochkarev, N.G., 1983, AZh 60, 448
\bibitem[1973]{bab1} Babadzhanyants M.K., et al., 1973, Trudy AO LGU 29, 72
\bibitem[1974]{bab2} Babadzhanyants M.K., et al., 1974, Trudy AO LGU 30, 69
\bibitem[1975]{bab3} Babadzhanyants M.K., et al., 1975, Trudy AO LGU 31, 100
\bibitem[1976]{bab4} Babadzhanyants M.K., et al., 1976, Trudy AO LGU 32, 52
\bibitem[1984]{bab5} Babadzhanyants M.K., et al., 1984, Trudy AO LGU 39, 43
\bibitem[1980]{bar1} Barr, P., et al., 1980, MNRAS 193, 562
\bibitem[1983]{bar2} Barr, P., Willis, A.J., Wilson, R., 1983, MNRAS 202, 453
\bibitem[1988]{baum} Baum, S.A., Heckman, T., Bridle, A., et al., 1988, ApJS 68, 643
\bibitem[1982]{blan} Blanforf, R.D., McKee, C.F., 1982, ApJ 255, 419
\bibitem[1982]{boch1} Bochkarev, N. G., Antokhin, I., 1982, Astron. Tsirk. 1238
\bibitem[1997a]{boch2} Bochkarev, N.G., Burenkov, A.N., and Shapovalova, A.I., 1997, Astron. \& Astroph. Trans. 14, 97
\bibitem[1997b]{boch3} Bochkarev N.G., Shapovalova A.I., Burenkov A.N., et al., 1997, Ap\&SS 252, 167
\bibitem[1999]{boch4} Bochkarev N.G., Shapovalova A.I., 1999, PASPC 175, 75
\bibitem[1971]{cannon} Cannon R.D., Penston M.V., and Brett R.A., 1971, MNRAS 152, 79
\bibitem[1994]{chakra} Chakrabarti, S.K., and Wiita, P., 1994, ApJ 434, 518
\bibitem[1989]{chen} Chen, K., Halpern, J.P., Fillipenko, A.V., 1989, ApJ 339, 742
\bibitem[2000]{chiang} Chiang, J., Reynolds, C.S., Blaes, O.M., et al. 2000, ApJ 528, 292
\bibitem[1987]{clav1} Clavel, J.C., Wamsteker, W., 1987, ApJ 320, L9
\bibitem[1991]{clav2} Clavel, J., et al., 1991, ApJ 366, 64
\bibitem[1998]{collier} Collier, S., et al., 1998, ApJ 500, 162
\bibitem[1999]{col1} Collin, S., Hur\'e\, J.M., 1999, A\&A 341, 385
\bibitem[2001]{col2} Collin, S., Hur\'e, J.M., 2001 
\bibitem[1976]{cousins} Cousins, A.W.J., 1976, Mem.R.Astron.Soc. 81, 25
\bibitem[1996]{crenshaw} Crenshaw, D.M., et al., 1996, ApJ 470, 322
\bibitem[1993]{die1} Dietrich, M., et al., 1993, ApJ 408, 416
\bibitem[1998]{die2} Dietrich, M., et al., 1998, ApJS 115, 185
\bibitem[2001]{doroshenko} Doroshenko, V.T., et al., 2001, Pis'ma v Astron. Zh. (in press)
\bibitem[1990a]{dum1} Dumont, A.-M. Collin-Souffrin, S., 1990a, A\&AS 83, 71
\bibitem[1990b]{dum2} Dumont, A.-M., Collin-Souffrin, S. 1990b, A\&A 229, 313
\bibitem[1998]{dum3} Dumont, A.-M., Collin-Souffrin S., Nazarova L., 1998, A\&A 331,11
\bibitem[1996]{edelson} Edelson, R.A., et al., 1996, ApJ 470, 364
\bibitem[1995]{erac1} Eracleous, M., Livio, M., Halpern, J.P., et al., 1995, ApJ 438, 610
\bibitem[1996]{erac2} Eracleous, M., Halpern, J.P., Livio, M., 1996, ApJ 459, 89
\bibitem[1997]{erac3} Eracleous, M., et al., 1997, ApJ 490, 216
\bibitem[1979]{gal} Galeev, A.A., Rosner, R., Vaiana, G.S., 1979, ApJ 229, 318
\bibitem[1983]{gas1} Gaskell, C.M. 1983, in: Proc. 24th Liege Int. Astrophys.Colloq., Quasars and Gravitational Lenses (Cointe-Ougree:Univ. Liege), 471
\bibitem[1986]{gas2} Gaskell, C.M., Sparke, L.S., 1986, ApJ 333, 646
\bibitem[1987]{gas3} Gaskell, C.M., Peterson, B.M., 1987, ApJS 65, 1
\bibitem[1996]{gas4} Gaskell, C.M., 1996, ApJ, 464 L107
\bibitem[1981]{gerbal} Gerbal, D., Pelat, D., 1981, A\&A 95,18
\bibitem[1981]{heckman} Heckman T.M., Miley, G.K., van Breugel, W.J.M., et al., 1981, ApJ 247, 403
\bibitem[2000]{hub} Hubeny, I., Agol, E., Blaes, O., et al., 2000, ApJ 533, 710
\bibitem[1994]{inda} Inda, M., et al., 1994, ApJ 420, 143
\bibitem[1994]{james} James, 1994, MINUIT : function minimization and error anaysis
(CERN program libr. long writeup D506)(Version 94.1; Geneva : CERN)
\bibitem[1966]{john} Johnson H.L., 1966, ARA\&A 4, 193
\bibitem[1996]{kaspi} Kaspi, S., et al., 1996, ApJ 470, 336
\bibitem[1995]{kor} Korista, K.T., et al., 1995, ApJS 75, 719
\bibitem[1989]{laor} Laor, A., Netzer, H., 1989, MNRAS 238, 897
\bibitem[1996]{law} Lawrence, C., et al., 1996, ApJS 107, 541
\bibitem[1991]{lea1} Leahy, J.P., Perley, R.A., 1991, AJ 102, 537
\bibitem[1995]{lea2} Leahy, J.P.,  Perley, R.A., 1995, MNRAS 277, 1097
\bibitem[1996]{lei1} Leighly, K.M., et al., 1996, ApJ 463, 158
\bibitem[1997]{lei2} Leighly, K.M., et al., 1997, ApJ 483, 767
\bibitem[1997]{lei3} Leighly, K.M. \& O'Brien, P.T., 1997, ApJ 481, L15
\bibitem[1997]{livio} Livio, M., Xu, C., 1997, ApJ 478, L63
\bibitem[1984]{lyuty} Lyuty V.M., Oknyansky V.L., and Chuvaev K.K., 1984, Pis'ma v Astron. 
Zh. 10, 803
\bibitem[1993]{maoz2} Maoz, D., et al., 1993, ApJ 404, 576
\bibitem[1986]{niez} Neizvestny, S.I., 1986, Bull. Spec. Astrophys. Obs. 51, 5
\bibitem[1982]{netzer} Netzer, H., 1982, MNRAS 198, 589
\bibitem[1990]{netz} Netzer, H., Maoz, D., 1990, ApJ 365, L5
\bibitem[2000]{nic} Nicastro, F., 2000, ApJ 530, L65
\bibitem[1998]{obri2} O'Brien, P.T., et al., 1998, ApJ 509, 163
\bibitem[1982]{orr} Orr, M.J.L., Browne, I.W.A., 1982, MNRAS  200, 1067
\bibitem[1975]{oster} Osterbrock, D.E., Koski, A.T., Phillips, M.M., 1975, ApJ 197, L41
\bibitem[1971]{pen1} Penston, M.J., Penston, M.V., Sandage, A., 1971, PASP 83, 783
\bibitem[1984]{pen2} Penston, M.V., Perez, E., 1984, MNRAS 211, 13P
\bibitem[1988]{perez} Perez, E., et al., 1988, MNRAS 230, 353
\bibitem[1983]{pet1} Peterson, B.M., Collins II, G.W., 1983, ApJ 270, 71
\bibitem[1991]{pet2} Peterson, B.M., et al., 1991, ApJ 368, 119
\bibitem[1993]{pet3} Peterson, B.M., 1993, PASP 105, 247
\bibitem[1993]{pet4} Peterson, B.M., et al., 1993, ApJ 402, 469
\bibitem[1994]{pet5} Peterson, B.M., et al., 1994, ApJ 425, 622
\bibitem[1995]{pet6} Peterson, B.M., et al., 1995, PASP 107, 579
\bibitem[1998]{pet6b} Peterson, B.M., et al., 1998, ApJ 501, 82
\bibitem[1999]{pet7} Peterson, B.M., et al., 1999, ApJ 510, 659
\bibitem[1980]{pica} Pica A.J., et al., 1980, AJ 85, 1442
\bibitem[1994]{reich} Reichert, G.A., et al., 1994, ApJ 425, 582
\bibitem[1990]{robi} Robinson, A., Perez, E., 1990, MNRAS 244, 138
\bibitem[1992]{rok1} Rokaki, E., Boisson, C., Collin-Souffrin, S., 1992, A\&A 253, 57
\bibitem[1997]{rodri} Rodriguez-Pascual, P.M., et al., 1997, ApJS 110, 9
\bibitem[1973]{sandage} Sandage A., 1973, ApJ 180, 687
\bibitem[1997]{santos} Santos-Lleo, M., et al., 1997, ApJS 112, 271
\bibitem[1976]{scott} Scott R.L., et al., 1976, AJ 81, 1440
\bibitem[1975]{selmes} Selmes, R.A., Tritton, K.P., Wordsworth R.W. 1975, MNRAS 170, 15
\bibitem[1973]{shak} Shakura, N. I., Sunyaev, R.A., 1973, A\&A 24, 337
\bibitem[1996]{shap} Shapovalova, A.I., Burenkov, A.N. and Bochkarev, N.G., 1996, Bull. Spec. Astrophys. Obs. 41, 28
\bibitem[1994]{stirpe} Stirpe, G.M., et al., 1994, ApJ 425, 609
\bibitem[2000]{ulr} Ulrich, M.H., 2000, ESO Prepr., N1359, 1
\bibitem[1991]{veilleux} Veilleux, S., Zheng, W., 1991, ApJ 377, 89
\bibitem[1993]{vlasyuk} Vlasyuk V.V., 1993, Bull. Spec. Astrophys. Obs. 36, 107
\bibitem[1981]{wam1} Wamsteker, W., 1981, A\&A 97, 329
\bibitem[1997]{wam2} Wamsteker, W., et al., 1997, MNRAS 288, 225
\bibitem[1999]{wand} Wandel, A., Peterson, B.M., Malkan, M.A., 1999 ApJ 526, 579
\bibitem[1997]{wan} Wanders, I., et al., 1997, ApJS 113, 69
\bibitem[1996]{war} Warwick, R., et al., 1996, ApJ 470, 349
\bibitem[1994]{whi} White, R.J., Peterson, B.M., 1994, PASP 106, 879
\bibitem[1998]{woz} Wozniak, P.R., et al., 1998, MNRAS 299, 449
\bibitem[1981]{yee} Yee, H.K., Oke, J.B., 1981, ApJ 248, 472
\bibitem[1991]{zhe1} Zheng, W., Veilleux, S., Grandi, S.A., 1991, ApJ 381, 121
\bibitem[1995]{zhe2} Zheng, W., et al., 1995, AJ 109, 2355
\bibitem[1996]{zhe3} Zheng, W., 1996, AJ 111, 1498
\end{thebibliography}
\end{document}